# Assessing the impact of soil moisture-temperature coupling on temperature extremes over the Indian region

Naresh G Ganeshi[1,2], Milind Mujumdar[1], Takaya Yuhei[3], Mangesh M Goswami[1], Bhupendra Bahadur Singh[1], R Krishnan[1], Toru Terao[4]

[1] Centre for Climate Change Research, Indian Institute of Tropical Meteorology, Pune

[2] Savitribai Phule Pune University, Pune

[3]Meteorological Research Institute, Climate Research Department, Ibaraki, Japan.

[4]Faculty of Education, Kagawa University, Kagawa, Japan.

**Data Availability Statement**

The control experiments i.e. historical and future 4K simulations from MRI-AGCM3.2 are free to download from d4PDF database http://search.diasjp.net/en/dataset/d4PDF_GCM and http://search.diasjp.net/en/dataset/d4PDF_RCM to all the users from August 2018. Temperature (minimum and maximum) and rainfall gridded data is available on IMD database https://imdpune.gov.in/Clim_Pred_LRF_New/Grided_Data_Download.html, both in binary and NetCDF format. The other datasets such as GLDAS is freely available to download from http://disc.gsfc.nasa.gov/hydrology, whereas, LDAS can be made available from IMD with the appropriate permission.

**Abstract**

Understanding the land-atmosphere interaction is one of the key aspects in the study of temperature extremes (ExT) over the tropical climate of India. While previous model sensitivity studies have mainly focused on discerning the soil moisture-precipitation feedback processes over the Indian region, the present study investigates the impact of soil moisture-temperature (SM-T) coupling on the ExT using the high-resolution (~60 km) model simulations. These simulations include the control and soil moisture (SM) sensitivity experiments (DRY-SM and WET-SM) initialized by perturbing (decreasing/increasing) SM from the historical (HIST: 1951-2010) and future 4K warming (FUT: 2051-2100) control runs. The analysis identifies the transitional regions of north-central India (NCI) as the hotspot of strong SM-T coupling. Over NCI, the HIST experiment shows an occurrence of 4-5 extreme events per year, with an average duration of 5-6 days per event and intensity exceeding 46$^{o}$C. Whereas, FUT estimates indicate relatively severe, long-lasting, and more frequent extreme events. The SM sensitivity experiments reveal the significant influence of SM-T coupling on the ExT over NCI in both historical and future climates. We find that the DRY-SM results in significant enhancement of frequency, duration and intensity of ExT, in contrast to WET-SM. This diagnosis is further supported by the Generalized Extreme Value (GEV) distribution method to carry out a quantitative assessment of the influence of SM on ExT over the NCI. We note that the difference between DRY-SM and WET-SM 50-year return value of the block maxima GEV fit can reach upto 1.25$^{o}$C and 3$^{o}$C for historical and future climate, respectively. The enhanced (reduced) extreme temperature conditions in DRY-SM (WET-SM) simulation are caused by the intensification (abridgement) of sensible heat flux by limiting (intensifying) available total energy for evaporative cooling due to faster (slower) dissipation of positive soil moisture anomalies (also called as soil moisture memory). In addition, the influence of SM on ExT over NCI is found to be larger during the post-monsoon season as compared to the pre-monsoon and monsoon seasons.

## 1. Introduction

Major land regions of the world have been exhibiting a severe rise in temperature extremes (ExT) during the past few decades (Perkins et al., 2012). Modelling studies for the future climate also reported a prominent increase in the extreme temperature characteristics all around the world (Brown, 2020; IPCC, 2018; Sanjay et al., 2020). The Indian subcontinent is highlighted as one of the hotspots for exacerbated ExT conditions during the last few decades (Perkins-Kirkpatrick and Lewis, 2020; Satyanarayana and Rao, 2020). Recent studies have projected that a spate of red-hot extremes is likely to become more common over the Indian region at the end of the 21st century (Das and Umamahesh, 2021; Krishnan et al., 2020; Mizuta et al., 2017; Murari et al., 2015; Rohini et al., 2019). Hot weather conditions over India generally persist during the pre-monsoon months of March, April, and May (De et al., 2005; Kothawale et al., 2010). These extreme heat conditions may extend during the Indian summer monsoon months as well as the post-monsoon season due to the prolonged monsoon break like conditions and land-surface relevance (Raghavan, 1966; Dimri, 2019; Ganeshi et al., 2020). With large spatio-temporal variability over the Indian region, warm extremes exert a serious impact on the ecosystem, human health, agriculture and economy (Im et al., 2017; Patz et al., 2005).

A primary cause for enhancing the frequency, duration and intensity of ExT is anticipated to the global warming, as a response to anthropogenic climate forcing (Baldwin et al., 2019; IPCC, 2021, 2014; Krishnan et al., 2020). The physical mechanism involved in the formation of such ExT is linked to large-scale atmospheric dynamics as well as regional-scale land-atmosphere interactions (De and Mukhopadhyay, 1998; Ganeshi et al., 2020; Ghatak et al., 2017; Joshi et al., 2020; Ratnam et al., 2016). The atmospherically driven phenomenon such as quasi-stationary Rossby wave-train and El Niño Southern Oscillation originated from the North-Atlantic Ocean and tropical Pacific, respectively, in-turn maintains the warm and dry air thermodynamically at the surface, through the high-pressure system. On the other hand, the land-atmosphere coupling can be a dominant aspect for explaining the processes underlying extreme heat magnitude, duration and severity (Fischer et al., 2007; Ganeshi et al., 2020; Gevaert et al., 2018).

Soil moisture (SM) has received a great deal of attention in weather and climate sciences. As a crucial component of the land-atmosphere interaction, SM also acts as temporal storage of atmospheric anomalies (Delworth and Manabe, 1988; Entin et al., 2000; Koster and Suarez, 2001; Wu and Dickinson, 2004a). This behaviour of the soil to persist the moist and dry conditions caused by atmospheric forcing is called soil moisture memory (SMM) or soil moisture persistence (Delworth and Manabe, 1988; Orth et al., 2013). This long-term persisting nature or memory of SM is associated with its low-frequency variability (Manabe and Delworth, 1990). Moreover, the low-frequency SM variability has the potential to induce the most pronounced impact on near-surface temperature and precipitation variability (D'Andrea et al., 2006). Therefore, as an essential variable of the climate system, the study of soil moisture variability provides key insights into the observational and modelling aspects of land-atmosphere feedback processes (Ganeshi et al., 2020; Miralles et al., 2012; Mujumdar et al., 2021; Seneviratne et al., 2010).

The influence of SM on near-surface temperature is mainly determined through limiting surface energy partitioning and evapotranspiration (Jaeger and Seneviratne, 2011). Several methods were proposed earlier to find the role of land-atmosphere coupling on ExT (Gevaert et al., 2018; Miralles et al., 2012; Seneviratne et al., 2010). One way of exploring the SM impact is by analysing composites of wet and dry land surface states associated with ExT (Rohini et al., 2016). The second approach uses the Generalized Extreme Value (GEV) theory for assessing changes in ExT, by fitting SM as covariate (Ganeshi et al., 2020; Whan et al., 2015). Such a stationary or non-stationary GEV method estimates return values of temperature maxima with or without considering SM as a covariate. The third way of understanding the SM impact is through performing the sensitivity experiment using the climate model run (Erdenebat and Sato, 2018; Seneviratne et al., 2006). These simulations allow us to investigate the one-way influence of SM on the atmosphere since SM feedback to the atmosphere is removed by controlling SM variability. Here, we utilize the novel method of SM sensitivity experiments over the Indian region to assess the role of wet and dry SM conditions on extreme temperature variability. The model experiments are carried out using the high-resolution (~60 km) version of MRI-AGCM3.2 (sub-section 2.1.1). Both wet and dry experiments are performed in this study as they depict the potential to reinforce the climate extremes (Liu et al., 2014).

Detailed understanding of causes for hot extremes and its increasing trend over India in past and future climate has been a topic of concern since the 1980s (Dimri, 2019; Im et al., 2017; Patz et al., 2005). The existing knowledge conveys that ExT occurrences over India have been discussed in the context of large-scale atmospheric dynamics and regional-scale land-atmosphere interactions, where the associated dry SM conditions are also highlighted as one of the contributing factors (Ratnam et al., 2016; Rohini et al., 2016; U S De and Mukhopadhyay, 1998). However, the impact of SM-T coupling on ExT is not exclusively understood due to insufficient spatio-temporal SM observations and, the difficulty in filtering the impact of climate extremes on SM with reanalysis and observational datasets. State-of-the-art model sensitivity experiments are being used to overcome the aforementioned issues and investigate the sensitivity of climate extremes to the SM variability over major areas of the world (Erdenebat and Sato, 2018; Seneviratne et al., 2006). In a review of SM sensitivity experiments over the Indian subcontinent, researchers mostly explored the soil moisture-precipitation feedback mechanism using the model simulations (Asharaf et al., 2012; Raman et al., 1998; Shukla and Mintz, 1982). However, in-depth investigation of SM-T coupling and its role on the ExT over the Indian region is still warranted. Therefore, the present study aims to unravel the underlying role of SM-T coupling on ExT by performing high-resolution (~60 km) SM sensitivity experiments from the MRI-AGCM3.2 model (refer to sub-section 2.1.1 for more details).

The outline of this paper is as follows: Section 2 presents the model simulation experiments, data and methodology used in this study. Section 3 includes a detailed description of strong land-atmosphere coupling regions, ExT, and the impact of SM on ExT over the Indian region analyzed using model simulations for historical (1951-2010) and future climate (2051-2100). Finally, section 4 summarizes the main findings of the study.

## 2. Data and methodology

### 2.1 Model simulation experiments

#### 2.1.1 Description of MRI-AGCM3.2 model

The revised version of the atmospheric general circulation model by Meteorological Research Institute (MRI-AGCM3.2; Mizuta et al., 2012) from its predecessor, MRI-AGCM3.1(Kitoh et al., 2009)is used in the present diagnostic study for analyzing ExT and land-atmosphere interaction over India. Here, we use the 60 km resolution version of MRI-AGCM3.2, having 64 vertical levels and 3 active soil layers. The model uses sea-surface temperature, sea ice concentration, and sea ice thickness as lower boundary conditions prescribed from Centennial Observation based estimation (COBE-SST2), Hirahara et al., 2014 and, Bourke and Garret, 1987, respectively. Whereas, the external forcing is configured with observed values of global mean concentration of greenhouse gases, MRI Chemistry climate model (MRI-CCM; Deushi and Shibata, 2011) output for three-dimensional distribution of ozone, and MRI Coupled Atmosphere-Ocean General Circulation Model (MRI-CGCM3; Yukimoto et al., 2011) output.

The dynamical framework of the MRI-AGCM3.2 version uses the hydrostatic primitive equation from the spectral transform method (Kanamitsu et al., 1983). For time integration, a two-time level semi-implicit Semi-Lagrangian scheme with the increase in computational stability is used in MRI-AGCM3.2 (Yoshimura and Takayuki, 2005). A new cumulus parameterization scheme (Yoshimura et al., 2015) is introduced in MRI-AGCM3.2 based on a scheme by Tiedtke (1989). The radiation scheme used in the model is similar to the Japan Meteorological Agency's operational model, except with the treatment of aerosols. MRI-AGCM3.2 version includes absorption due to water vapour (line and continuum absorption), carbon dioxide (in the 15 mm band, near-infrared region, etc.), and ozone (in the 9.6 mm band, visible and ultraviolet region). The model also deals with absorption due to methane ($CH_4$), dinitrogen monoxide ($N_2O$), and chlorofluorocarbons (CFCs) in the longwave scheme, for considering their greenhouse effect. Whereas, shortwave radiation includes the absorption due to oxygen and Rayleigh scattering by molecules of atmospheric gas. The optical parameters are configured to five types of aerosol species: sulfate, black carbon, organic carbon, mineral dust, and sea-salt while considering the direct effect of aerosol. All computations related to absorption coefficient and asymmetry factors for aerosol species in the model are based on an assumption of Mie scattering by spherically shaped particles. MRI-AGCM3.2 implemented with the improved version of Simple Biosphere Model as land surface scheme to estimate the temporal changes in various land-surface properties like soil, snow, surface skin and vegetation (Hirai et al., 2007). For the boundary level mixing scheme, the Mellor-Yamada level 2 turbulence closure scheme (Mellor and Yamada, 1974) is implemented in a model. For more details of MRI-AGCM3.2, please refer to Mizuta et al., (2012).

MRI-AGCM3.2 is extensively used in the investigation of numerous studies such as extreme events, simulating global precipitation and its change in future climate, and reproduction of tropical cyclones (Kusunoki, 2017; Mizuta et al., 2012). Furthermore, model validation with IMD observations and various assimilation products illustrate the least bias in the simulations corresponding to the SM-T coupling characteristics (described in subsection 3.1). As the model biases are not critical, the study further performs the soil moisture sensitivity experiments to understand the impact of SM on ExT.

#### 2.1.2 Experimental setups

In the present study, we have conducted six long-term high-resolution MRI-AGCM3.2 climate simulations (listed in table 1): 1) historical run (HIST: 1951-2010), 2) historical-20%SM (HIST-20: 1951-2010), 3) historical+20%SM (HIST+20: 1951-2100), 4) future 4K projection (FUT: 2051-2100), 5) future-20%SM (FUT-20: 2051-2100) and 6) future+20%SM (FUT+20: 2051-2100). The experimental setup of HIST and FUT are the same as those of d4PDF (Mizuta et al., 2017). Historical simulation experiments use both natural (e.g. volcanoes and solar variability) and anthropogenic forcing (e.g. greenhouse gases (GHG), aerosols etc.). Whereas, future 4K experiments identify a future climate in which global mean temperature becomes 4K warmer than pre-industrial climate. These FUT simulations use the detrended observed SST pattern with added climatological warming SST measure for experiments**.** In addition to historical and future 4K simulations (HIST and FUT), we have performed wet and dry control soil moisture sensitivity experiments (HIST-20, HIST+20, FUT-20 and FUT+20) for each of two long term simulations (HIST and FUT) to evaluate the role of SM on ExT over the Indian region. The simulations are initialized on the 1$^{st}$ day of each month by perturbing top layer SM (upto 10 cm depth) initial conditions with corresponding fields from HIST and FUT simulations. WET-SM simulations are initialized by setting up the 20% increased SM fields, whereas in DRY-SM experiment surface SM is set to 20% decreased conditions at each grid point. Both DRY-SM and WET-SM sensitivity experiments were initialized with the same initial lateral boundary conditions as that of HIST and FUT simulations except for the above-mentioned SM perturbation. The impact of SM on ExT is quantified by analyzing the WET-SM and DRY-SM experiments.

**Table 1.** List of experiments

| **Experiment** | **Forcing used** | **Period** |
|---|---|---|
| HIST | Natural (e.g. volcanoes and solar variability) and anthropogenic forcing (e.g. greenhouse gases (GHG), aerosols etc.). | 1951-2010 (60 years) |
| FUT | Future climate in which global mean temperature becomes 4K warmer than pre-industrial climate. SST prescribed from detrended observed value with the addition of climatological warming pattern (Mizuta et al., 2017) | 2051-2100 (50 years) |
| DRY-SM (HIST-20) | The sensitivity experiment initialized by decreasing the soil moisture on 1$^{st}$ day of each month by 20% in HIST. | 1951-2010 (60 years) |
| DRY-SM (FUT-20) | The sensitivity experiment initialized by decreasing the soil moisture on 1$^{st}$ day of each month by 20% in FUT. | 2051-2100 (50 years) |
| WET-SM (HIST+20) | The sensitivity experiment initialized by increasing the soil moisture on 1$^{st}$ of each month by 20% in HIST and FUT. | 1951-2010 (60 years) |
| WET-SM (FUT+20) | The sensitivity experiment initialized by increasing the soil moisture on 1$^{st}$ of each month by 20% in FUT. | 2051-2100 (50 years) |

### 2.2 Other data products

With the focus on the ExT and land-atmosphere interaction, we have validated and corrected the model simulations using the observational and data assimilation products over the Indian region. Observational data include mean temperature, maximum temperature and precipitation data sets prepared by India Meteorological Department (Pai et al., 2015; Srivastava et al., 2009). Two data assimilation products are used in the present study: 1) Global Land Data Assimilation System (GLDAS), and 2) Land Data Assimilation System (LDAS) to validate and remove the bias in model outputs. The GLDAS data at resolution 0.25°×0.25° is used here to examine the evapotranspiration (ET), and surface energy fluxes (sensible and latent heat fluxes) over the Indian region (Rodell et al., 2004). The National Aeronautics and Space Administration (NASA) Goddard Space Flight Center (GSFC) and the National Oceanic and Atmospheric Administration (NOAA) National Centers for Environmental Prediction (NCEP) jointly contributed for the preparation of GLDAS data. GLDAS data is generated by combining satellite and ground-based observational products into land surface modelling and data assimilation techniques. Previous research has shown that GLDAS outputs are in good agreement with observations over the Indian region(Ganeshi et al., 2020; Mujumdar et al., 2021b; Sathyanadh et al., 2016). Out of four land-surface models driven by GLDAS, the present study uses the Noah land Surface Model version3.3 (Noah LSM 3.3) forced by the global meteorological forcing dataset (Sheffield et al., 2006). Furthermore, the high resolution (~ 4 km) data generated by using LDAS (version 3.4.1) is explored here to validate and remove the bias in the SM over the Indian region (Nayak et al., 2018). The LDAS version is mainly designed for both coupled and uncoupled modes within the Weather

Research and Forecasting model. More details of the LDAS high-resolution SM data product can be found in Nayak et al. (2018).

### 2.3 Analysis methodology

#### 2.3.1 Soil moisture-temperature (SM-T) coupling

Determining the SM-T coupling is an important factor for assessing the impact of SM on ExT (Seneviratne et al., 2006). Based on the model experiments and analytical techniques several methods were discussed earlier to evaluate the coupling strength between SM and surface temperature (Dong and Crow, 2018; Miralles et al., 2012; Seneviratne et al., 2013). In the present study, quantification of surface temperature sensitivity to the SM change (SM-T coupling strength) has been carried out using the method suggested by Dirmeyer (2011). Here, we are using a similar method for surface temperature sensitivity instead of surface energy fluxes. The method described by Dirmeyer (2011) overcomes the shortcomings in the correlation method by considering the variance of SM at each grid point. SM-T coupling metric used in this study is given in equation 1. In this method, we estimate the linear regression slope ($\mathcal{R}_c$) of temperature anomaly on the SM anomaly. Furthermore, the coupling metric is determined by multiplying the negative value of soil moisture standard deviation ($\sigma_{SM}$) to regression slope ($\mathcal{R}_c$).

$$\text{Coupling strength}(\Omega) = -\mathcal{R}_c * \sigma_{SM} \quad (1)$$

Whereas, $\mathcal{R}_c$ denotes the slope of linear regression of temperature anomaly on SM anomaly and $\sigma_{SM}$ indicates the standard deviation of soil moisture at each grid point. Higher positive values of $\Omega$ indicate the regions where SM has a dominant impact on temperature variability. Whereas, lower values point towards the region with the insignificant impact of SM on surface temperature and thus on extremes.

### 2.3.2 Extreme temperature indices

Extreme temperature conditions can be detected using various criteria depending upon the different climate zones (Nairn and Fawcett, 2013). The most common definitions of temperature extremes mainly depend upon the daily maximum temperature for the analysis. The present study uses three different indices based on the daily maximum temperature from the MRI-AGCM3.2 model for the evaluation of ExT characteristics over India. The indices used here are similar to the standard ExT indices as referred to by ETCCDI (Roy, 2019). Moreover, considering the severity of ExT beyond the pre-monsoon season, the present study evaluates the extreme temperature characteristics on an annual scale (Ganeshi et al., 2020; Ramarao et al., 2016). For the historical climate, extreme temperature event is defined if the daily $T_{max}$ value at each grid point is at least 3$^{o}$C greater than the 90$^{th}$ percentile $T_{max}$ of the corresponding day and persists at least for three consecutive days. Future climate uses the same definition for detecting ExT with the 90$^{th}$ percentile threshold from the HIST experiment. Furthermore, the total number of extreme temperature events per year is considered for defining the first index i.e. Extreme Temperature Frequency (ExTF) and the total number of days in each event is counted as the Extreme Temperature Duration (ExTD) index. The third index i.e. Extreme Temperature Intensity (ExTI) is the measure of maximum $T_{max}$ for each year at each grid point.

### 2.3.3 Generalized Extreme Value (GEV) distribution

The quantitative description of the impact of SM on ExT is carried out in the present study using the statistical approach of GEV theory (Whan et al., 2015). Block maxima perspective is demonstrated to fit GEV to the yearly maximum temperature (ExTI) at each grid point, with and without considering SM as a covariate. The extReme software package of R-programming to perform the GEV fit is used in this study (R Core Team, 2016; Gilleland and Katz, 2016). The GEV analysis is carried out in two important steps. The first step of analysis consists of stationary GEV fit for block maxima of each year with no covariate (given below in equation 2). Whereas, the second step consists of a non-stationary GEV model fit to ExTI with the inclusion of SM as a covariate. The GEV analysis is dependent upon the three important parameters: 1) scale parameter ($\sigma$), 2) location parameter (μ), and 3) shape parameter (ξ). Scale parameter ($\sigma$) explains the variability in the dataset, mean of the fitted distribution is represented in terms of location parameter(μ) and the shape of the fitted distribution is shape parameter (ξ) (i. e. Gumbel:$\xi = 0$, Frechet:$\xi > 0$, Weibull:$\xi < 0$). The GEV analysis is carried out on the area-averaged value of ExTI and SM over the hotspot of strong land-atmosphere coupling (NCI).

$$G(x) = \exp\left[-\left(1 + \frac{\xi(x-\mu)}{\sigma}\right)\right], \quad (2)$$

$$\mu(y) = A_0 + A_1\, y, \quad (3)$$

where, y is standardized annual mean SM and, $A_0$ and $A_1$ are fitting constants for location parameters.

The negative value of the estimated shape parameter for the HIST and FUT experiment indicates the ExTI follows the Weibull distribution (Table 2 and 4). The significance for perfect stationery and non-stationary GEV fit using the likelihood-ratio-test (LRT) showed a good fit for ExTI distribution and inclusion of SM as covariate dominantly improves the model fit. In the GEV analysis impact of SM on extremes is documented by analyzing differences between 50-year return levels of ExTI from DRY-SM and WET-SM sensitivity experiments for historical and future climate.

**Table 2** Statistics for stationary GEV model fit (HIST)

| **Estimated parameters** | **Lower 95 % confidence interval** | **Estimates** | **Lower 95 % confidence interval** | **P-value** |
|---|---|---|---|---|
| Location (μ) | 45.69 | 45.06 | 46.44 | 0.05 |
| Scale($\sigma$) | 1.07 | 1.37 | 1.67 | |
| Shape(ξ) | -0.73 | -0.57 | -0.41 | |

**Table 3** Statistics for non-stationary GEV model fit with SM as covariate (HIST)

| **Estimated parameters** | **Lower 95 % confidence interval** | **Estimates** | **Lower 95 % confidence interval** | **P-value** |
|---|---|---|---|---|
| $A_0$ | 47.55 | 53.62 | 57.69 | 0.05 |
| $A_1$ | -53.67 | -30.19 | -6.69 | |
| Scale($\sigma$) | 1.003 | 1.307 | 1.61 | |
| Shape(ξ) | -0.80 | -0.59 | -0.37 | |

**Table 4** Statistics for stationary GEV model fit (FUT)

| **Estimated parameters** | **Lower 95 % confidence interval** | **Estimates** | **Lower 95 % confidence interval** | **P-value** |
|---|---|---|---|---|
| Location (μ) | 48.94 | 49.69 | 50.43 | 0.05 |
| Scale ($\sigma$) | 1.09 | 1.64 | 2.2 | |
| Shape (ξ) | -0.7177 | -0.462 | -0.2073 | |

**Table 5** Statistics for non-stationary GEV model fit with SM as covariate (FUT)

| **Estimated parameters** | **Lower 95 % confidence interval** | **Estimates** | **Lower 95 % confidence interval** | **P-value** |
|---|---|---|---|---|
| $A_0$ | 50.93 | 58 | 65.07 | 0.05 |
| $A_1$ | -66.38 | -35.19 | -4.009 | |
| Scale ($\sigma$) | 0.875 | 1.449 | 2.1167 | |
| Shape (ξ) | -1.074 | -0.03 | -0.2017 | |

### 2.3.4 Soil moisture memory (SMM)

Property of the soil to remember wet or dry anomalies caused by atmospheric forcing is generally termed as soil moisture memory (Delworth and Manabe, 1988; Wu and Dickinson, 2004b). The present study measures the SMM in terms of time-scale lag at which the autocorrelation drops to 1/e (e-folding time scale) of its value (Ganeshi et al., 2020). The method of e-folding time scale is based on the 30-day lag autocorrelation values of soil moisture anomalies considering the exponential decay of SM autocorrelation function (equation 4).

$$r(\tau) = e^{-\tau/\lambda} \qquad \ldots (4)$$

where r (τ) is the autocorrelation function, τ is the lag and λ is called decay time-scale. SMM analysis is also extended to the WET-SM and DRY-SM experiments to explore the impact of SMM on ExT.

## 3. Results and discussion

### 3.1 Validation of model simulations with observational and assimilation data sets

The present study uses three different types of data sets (see section 2.2) to validate six variables (precipitation: PR, soil moisture: SM, maximum temperature: $T_{max}$, evapotranspiration: ET, sensible heat flux: SHF, and latent heat flux: LHF) from the MRI-AGCM3.2 historical experiment. Fig. 1 shows the spatial distribution of uncorrected fields (column I), mean bias (column II) and corrected model outputs (column III) from MRI-AGCM3.2 (spatial distribution is based on long-term mean i.e. 60-year average climatology). A more detailed description of IMD gridded observations and data assimilation products (LDAS and GLDAS) used to evaluate model simulations are discussed in section 2.2. Overall, MRI-AGCM3.2 well captures climatological features of all the variables (PR, $T_{max}$, SM, SHF, LHF and ET) used in the study. Although some relatively large inconsistency in the spatial distribution of model biases among all outputs, by and large, maximum bias can be observed over the north, north-east and along foot-hills of Himalaya in simulated variables.

More accurate representation of precipitation using model simulation always remains a challenging point in model development. Fig. 1a and m show the mean spatial pattern of precipitation (uncorrected and corrected) from MRI-AGCM3.2 over the Indian region. Results indicate that MRI-AGCM3.2 reasonably captures observed climatological PR distribution over the Indian region. The Western Ghats and north-east Indian regions are indicated to have the highest mean PR exceeding 7 to 8 mm/day with an average bias ranging from -2 to 2 mm/day over the Western Ghats and -3 to -1 mm/day over north-east India. On the other hand, the least PR (1 mm to 2 mm/day) can be observed over the north-west part of India, which is mainly considered as the arid region (Ramarao et al., 2018). The north-west Indian region is also showing less positive bias in the model PR as compared to the IMD observation (less than 1 mm/day). Furthermore, the highest mean dry bias in model PR can be observed over the northern part of the Indian region with a bias value less than -3 mm/day. In accordance with the IMD observation, MRI-AGCM3.2 indicates moderate PR conditions (3 to 6 mm/day) over the central Indian region, with bias ranging from -2 to 2 mm/day.

Climatology of long-term annual average daily corrected maximum temperature ($T_{max}$) is shown in Fig. 1n. The spatial pattern indicates that MRI-AGCM3.2 well captures $T_{max}$ distribution over the Indian region. Overall, the model shows the cold bias in $T_{max}$ across the Indian region. $T_{max}$ distribution in MRI-AGCM3.2 mostly depends on the latitudinal variation, indicating warmer temperature conditions over southern peninsular India than the northern latitudes. Except for the north-east and north Indian regions, the annual average maximum temperature over India varies between 24 °C to 32 °C. Whereas, north and north-east regions are relatively cooler ($T_{max}$ < 20 °C) than the interior parts of India. An important point to be noted from Fig. 1h is that the model is having maximum cold bias over the north (bias < -5 °C) and north-east (bias< -2 °C) Indian region. However, relatively less cold bias over the north-central and southern peninsular India shows the reliability of model data sets over the regions.

The present study uses LDAS data set to validate the SM model output. The spatial distribution of SM over India indicates a direct association with PR. Wet SM conditions are mostly observed over the Western Ghats, north-east India and north Indian region (Fig. 1o). The wet regions except north India indicate very less bias of about 10% (-0.03 to 0.03 $m^3/m^3$) of mean SM conditions. Whereas, the north Indian region shows nearly 25% (bias< -0.09 $m^3/m^3$) bias in the model. Moderate dry bias can be observed over west-central India, Interior Maharashtra, Interior Karnataka and Tamilnadu (-0.06<bias<-0.03). Consistent with the rainfall conditions, MRI-AGCM3.2 indicates moderate SM over the central Indian region. North-west Indian region shows to have drier SM conditions, with a bias range of -0.01 to 0.01 $m^3/m^3$.

Model simulated fluxes and evapotranspiration (ET) are evaluated here with the help of the GLDAS product. ET is a crucial element in the hydrological cycle and this process is mainly controlled by the amount of water available at the top layer of the soil, vegetation and radiative flux (Fig. r). On average, MRI-AGCM3.2 underestimates ET over the Indian region (Fig. 1 l). The Western Ghats, north-central, north and north-east Indian regions indicate the larger negative bias of -1 to -2 mm/day. Model simulation mostly shows larger values of ET over the transitional region where a sufficient amount of SM strongly constrains ET variability due to available radiational energy (Fig. 1r). In agreement with a previous study by Ganeshi et al. (2020), the transitional regions are mostly located over central India. On the other hand, less ET over the north-west and north Indian region is observed due to the SM and energy limiting conditions, respectively. Describing other two fluxes, in which latent heat flux (LHF) is closely associated with the ET process and the SHF has the dominant influence from radiational energy (Fig. 1 p and q). Spatial distribution of biases SHF and LHF showing the contrast behaviour with respect to each other (Fig.1 j & k).

The present study uses a monthly bias-corrected method by Soriano et al. (2019) to correct bias in the model outputs. Bias corrected climatological features of all the variables are shown in Fig. 1. The result indicates that the method is able to reduce the bias induced in the model. Comparison of bias-corrected fields from MRI-AGCM3.2 with observational and reanalysis data sets is carried out with the help of Taylor statistics (Taylor, 2001). Three statistical scores (correlation, standard deviation and root mean square error) are used for the Taylor method to analyze corrected outputs (Fig. 2). All the corrected model outputs are standardized and averaged over the Indian region before representing through the Taylor statistics. The result shows a significant improvement (with correlation $> 0.7$ and RMSE $< 1$) in the model performance using the monthly correction method of Soriano et al. (2019) (Fig. 2). It can be observed that SM and SHF have the highest correlation values greater than 0.9 and least RMSE ($< 0.5$) with respect to LDAS and GLDAS, respectively. Whereas, the lowest correlation (0.7 to 0.8) is observed in the PR and LHF with RMSE close to 0.65.

### 3.2 Mean features of hydro-meteorological variables for future simulation

This section describes the mean features of six different hydro-meteorological variables (PR, $T_{max}$, RF, SHF, LHF and ET) over India for future climate (FUT) and its change with respect to the historical simulations (HIST). FUT simulation indicates a similar spatial distribution of all hydro-meteorological variables as that of HIST simulations (Fig. 1 and 3) with moderate change in the magnitude. Spatial distribution of PR over India in FUT simulation represents the highest mean value over the Western Ghats, north-east India and foot-hills of Himalaya exceeding the 9 mm /day. Whereas, least PR in future can be observed over north-west India and, the interior of Maharashtra and Karnataka (PR < 4 mm/ day). Overall, Fig. 3b shows that PR is projected to increase almost everywhere over the Indian region under the 4K warming scenario. The highest change (increase) in the FUT PR will likely to be observed over the large area of Maharashtra, Karnataka and north-east Indian region (PR >1 mm/day) with respect to HIST simulation. In contrast to PR distribution, model indicates the rise in maximum daily temperature ($T_{max}$) over the Indian region. With the 4K warming scenario, $T_{max}$ over the Indian region will likely to be increased minimum by 2$^{o}$C. The change in $T_{max}$ is likely to reach upto 3 $^{o}$C over the north-west, south and north Indian region in FUT climate relative to HIST simulation. In the global context, the fifth assessment report by IPCC concluded a similar result based on representative concentration pathways. AR5 assessment documented the rise in global mean surface temperature and precipitation at the end of the twenty-first century under RCP scenarios (IPCC, 2014).

Climatological spatial distribution of SM over the Indian region in future climate indicating to have a direct association with PR (Fig. 3c). With an increase in PR in future climate, the model indicates an increase of SM almost everywhere over the Indian region. FUT model simulation shows that, with an average 1 mm/day increase in PR over the Indian region, SM will likely to be increased by at least 2% as compared to HIST simulations. Wet conditions (increase in PR and SM) and warming over the Indian landmass are expected to reinforce the surface energy partitioning and ET rate in the future climate. The model result indicates that an increase in SM and temperature will be expected to lose more energy to the atmosphere in terms of LHF, through the increase in the process of ET (Fig. 3j and l). Unlike the LHF, overall, FUT simulation resulted in less amount of heat transfer to the atmosphere by the sensible heating process as the dominance of energy released accounted during the latent cooling process (Fig.3h). The relation between SM, $T_{max}$, SHF and LHF is highly non-linear, which can be defined using the process of land-atmosphere coupling (Ganeshi et al., 2020; Miralles et al., 2012; Seneviratne et al., 2013).

### 3.2 Soil moisture-temperature (SM-T) coupling over the Indian region

Soil moisture has a dominant influence on temperature variability and, as a result, on ExT over the regions of strong land-atmosphere coupling (Seneviratne et al., 2010). Here, we aim to understand the SM-T coupling in historical (1951-2010) and future (2051-2100) climate through the linear regression method (sub-section 2.3.1; Dirmeyer, 2011). In addition, diagnosis of coupling strength extended on the entire annual cycle to explore the role of SM on annual extremes beyond pre-monsoon months. Fig. 4 indicates the spatial distribution of SM-T coupling strength (Ω) across the Indian region. The result shows that hotspot of strong soil moisture-temperature coupling is located over the north-central Indian (NCI) region. Stronger coupling over the NCI reveals the significant control of land-atmosphere interaction on near-surface temperature. The spatial pattern of SM-T coupling nearly coincides with the coupling hotspot highlighted in a recent study by Ganeshi et al (2020) over India. The coupling strength from model simulations is also verified with the recently developed coupling method by Miralles et al. (2012). It can be observed from the supplementary Fig. S1 and Fig. 4a that the spatial coupling pattern from metric π and Ω are consistent with each other.

SM-T coupling over the landmass is mainly limited by the combined effect of water availability at the top surface and radiation energy. Lower coupling strength over the wet and dry regions is due to the shortage of energy availability and less evaporation variability, respectively. On the contrary, imposing dominant control on evaporation variability, moderate SM regimes of NCI indicate to have a larger impact on near-surface temperature variability (Ganeshi et al., 2020; Seneviratne et al., 2010). Investigation of SM-T coupling further extended for the future climate under the 4K warming scenario. By and large, the results from future climate simulations present a similar spatial distribution of SM-T coupling over India as that of historical climate. Following the results from historical climate, the coupling index underlines the NCI region as a hotspot of strong soil moisture-temperature interaction, with an increase in coupling magnitude (Fig. 4b). From Fig. 4a and b, it is to be noted that the area of strong SM-T coupling is likely to expand under future warming scenarios. The expansion or shrinking of strong soil moisture-temperature coupling regions can be an important factor in the backdrop of climate change (Krishnan et al., 2020).

### 3.3 Long-term mean of temperature extremes

Figure 5 shows the spatial distribution of the long-term mean extreme frequency (ExTF), duration (ExTD) and intensity (ExTI) over the Indian region for historical (1951-2010) and future (2051-2100) climate. Estimates of ExTF, ExTD and ExTI are carried out using bias-corrected $T_{max}$ simulations from MRI-AGCM3.2. Spatial maps of extremes for historical climate indicate at least four events per year over the Indian landmass with an average duration of ~ 5 to 6 days per event (Fig. 5). In addition, the intensity of extremes is found to be maximum over the central Indian region (>47$^{o}$C) with a pattern correlation of about 0.9 (significant at 95 % confidence) with respect to IMD data. Verification results for ExTD, ExTF and ExTI are showing good agreement of historical model $T_{max}$ simulations with the IMD gridded data set. Pattern correlation of ExTD,

ExTF and ExTI from MRI-AGCM3.2 with IMD is greater than 0.4 at 90% confidence. Future changes in extremes for the period 2051-2100 are also evaluated here under the 4K warming scenario. Future simulation suggests the immense increase in extreme temperature characteristics over the Indian region under the 4K warming scenario. Our finding shows that high intensity (ExTI> 50 $^{o}$C) extreme temperature events are likely to occur after every 25-30 days in future (4K warming scenario) over India, which can prevail at least for 2 to 4 days.

Figure 5 d, h and l show the spatial difference of future extreme temperature indices with respect to HIST. Under the 4K warming scenario, extreme temperature cases in future climate were likely to be two times higher than that of historical climate. On average, the model result indicates an increase of about ~9 ExTF per year, 5-6 ExTD per event and 3$^{o}$C ExTI in future climate as compared to historical climate. The multiple order higher difference with respect to historical climate indicates the severity extremes in future climate. We further carried out the analysis of extremes over the strong soil moisture-temperature coupling regions of the NCI, in order to understand the SM impact on temperature extremes. Area-averaged time series over the NCI is used to evaluate the long term changes in ExTD, ExTF and ExTI for the historical (1951-2010) and future climate (2051-2100). Over NCI, historical model simulation estimates 4-5 extreme episodes per year, with an average intensity greater than 46$^{o}$C and duration of 5-6 days per occurrence (supplementary Fig.S2). Extremes are even deadlier in future climate under 4K warming scenarios. Future projections demonstrate very intense (ExTI> 50 $^{o}$C), long persisting (ExTD> 10 days per event) and more frequent (ExTF>15 events per year) extreme events as those of historical climate under 4K warming scenario (Supplementary Fig. S3).

### 3.4 Impact of soil moisture on temperature extremes in historical and future climate

In the present study, the influence of soil moisture on ExT is diagnosed using sensitivity experiments (WET-SM and DRY-SM experiments). These experiments are carried out for historical and future climates over the Indian region (see sub-section 2.1.2). In the WET-SM experiment, SM at each grid point is increased by 20% on the 1$^{st}$day of each month. Whereas, for the DRY-SM experiment, SM is decreased by 20% on the 1$^{st}$day of each month at each grid point. Fig. 6 shows the annual climatology for ExT diagnosis from HIST, HIST+20 and HIST-20 experiments (described in the sub-section 2.1). On average, drier SM conditions (HIST-20) can increase the ExTF by 4-5 events per year, ExTD by 1-2 days per event and long-term mean ExTI > 0.6 $^{o}$C as compared to HIST (Fig. 6 b, e& h). In contrast, HIST+20 simulations tend to reduce ExT over the Indian region. Hist+20 sensitivity experiment shows an average decrease of frequency (ExTF), duration (ExTD) and intensity (ExTI) by 1-2 events/year, 2-3 days per event and ~ 0.5$^{o}$C (long-term mean), respectively, as that of HIST (Fig. 6 c, f & i). Future climate sensitivity experiments demonstrate similar results to historical simulations, albeit with a smaller impact of soil moisture. FUT-20 simulation leads to intensifying the extremes by 1-2 events per year (ExTF), 0-1 days per event and long-term mean intensity ~1 $^{o}$C than that of FUT (Fig. 7 b, e & h). Unlike the HIST+20 experiment, the FUT+20 simulation exerted a higher impact of wet SM for reducing ExT over the Indian region. FUT+20 result indicates that wet conditions in future reduce the ExTF by 3-4 events per decade, ExTD by 3-4 days per event and long-term mean (for period 2051-2100) intensity by ~2$^{o}$C (Fig. 7 c, f &i).

The main aim of this study is to understand the role of SM on ExT over the hotspot of SM-T coupling. Therefore, further sensitivity analysis is carried out over the strong SM-T coupling regions of north-central India (NCI). Historical results over the NCI suggest that DRY-SM (HIST-

20) leads to an increase in the ExTF ~5 events per year, ExTD ~ 1.8 days per event and long-term mean ExTI ~ 0.71$^{o}$C, as compared to HIST (Fig. 8). Whereas, wet simulation (HIST+20) reduces the ExTF ~3 events per year, ExTD ~1 day per event and long-term mean ExTI ~ 1.88 $^{o}$C as compared to HIST. Future experiments for SM impact over NCI shows the increase of 2.2 events per year, 1.55 days per event and long-term mean intensity ~ 0.93$^{o}$C under dry SM conditions (FUT-20) as that of future climate. While, in the case of wet simulations (FUT+20) significant decrease in the extreme temperature characteristics (decrease of ExTF ~ 3.3 events per year, ExTD by 2 days per event and long-term mean ExTI ~ 2.02$^{o}$C) is observed over the NCI. The sensitivity experiments revealed the significant impact of SM on ExT characteristics over the Indian region. Moreover, the dominant influence of SM on ExT can be found over the hotspot of strong SM-T coupling.

Analysis of extremes over the NCI is also supported by the Generalized Extreme Value (GEV) theory and probability distribution approach. The yearly block maxima approach is applied to the non-stationary GEV model fitting of extreme temperature intensity (ExTI) index considering soil moisture as a covariate. Further, the SM influence on extremes is quantified using the difference between 50-years return values of DRY-SM and WET-SM sensitivity experiments. Fig. 9 shows the return level plot for yearly block maxima fit to non-stationary GEV model in the case of WET-SM (Hist+20/Fut+20) and DRY-SM (Hist-20/Fut-20) experiments. Fig. 9a and b indicates the higher return level values of yearly $T_{max}$ in DRY-SM simulation (red colour curve) than the WET-SM experiment (blue colour curve). For the historical period, the difference between 50-year return values of DRY-SM and WET-SM simulation attains nearly ~ 1.25 $^{o}$C. In other words, a decrease of SM by 20% over the NCI lead to extend the yearly maximum temperature upto 48.75 $^{o}$C once in 50-years. On the other hand, in the case of wet simulation (HIST+20), the yearly max temperature remained below 47.63 $^{o}$C once in 50-year. Furthermore, DRY-SM/FUT-20 (WET-SM/FUT+20) for the future climate is predicting an intensification (reduction) of yearly maximum temperature upto (below) to 53.52$^{o}$C (50.94 $^{o}$C) for a 50-year return period.

To strengthen the analysis of ExT, furthermore, the influence of SM on extremes over NCI is discussed here using the PDFs of yearly block maxima (ExTI) for control (HIST/FUT) and DRY-SM sensitivity simulation (HIST-20/FUT-20). The quantitative investigation is done by calculating the probability of extremes exceeding the 90$^{th}$ percentile of PDF value. Fig.10 shows the probability distribution of extremes for dry (HIST-20/FUT-20) and control (HIST/FUT) simulation. Both the PDFs for DRY-SM experiments are indicating the significant shift towards the higher end of the distribution. For historical simulation, a decrease of SM by 20% (HIST-20) over the NCI enhances the 90$^{th}$ percentile probability limit of yearly block maxima of ExTI by 0.5$^{o}$C. On the other hand, drier soil moisture conditions can amplify the 90$^{th}$ percentile probability of extremes by 0.74 $^{o}$C in future climates. The model simulation also revealed the increase in the impact of SM on ExTI in the future climate under the 4K warming scenario over NCI. The possible reason for increasing SM influence over NCI could be linked with wetning or drying of land surface states, however, further research needs to be carried out to evaluate such effect.

The present study also evaluates the seasonal contribution of SM on ExT using PDFs of seasonal maxima over the NCI. Three seasons i.e. pre-monsoon (March to May), monsoon (June to September) and post-monsoon (October to December) were evaluated here to diagnose the SM impact. Results for both the historical and future simulations illustrate the significant impact of drier SM conditions in all three seasons. Sensitivity experiments prominently brought out that SM has a maximum impact on extremes in the post-monsoon season than that of pre-monsoon and monsoon seasons (Fig. 11 and Fig. 12). For historical climate, the probability of extremes with intensity greater than the 90th percentile of PDF distribution is increased by a minimum of 0.5 $^{o}$C (Fig. 11). In DRY-SM simulation (HIST-20) during the post-monsoon season, the probability of ExTI over NCI is increased by nearly 3.5 $^{o}$C compared to the control run (HIST). However, the likelihood of future post-monsoonal ExTI over NCI is expected to rise by at least 4$^{o}$C (Fig. 12). In summary, analysis of sensitivity experiments along with the return values of GEV distribution fit and PDF distribution revealed the crucial role of SM on ExT over the regions of strong SM-T coupling. Therefore, the accurate projections of the future soil moisture, rainfall and model representation of soil moisture-temperature coupling are important factors for making accurate projections of temperature extremes. The processes illustrating the influence of SM on extremes through land-atmosphere coupling is discussed in the next sub-section using the sensitivity experiments.

### 3.5 Factors governing the impact of soil moisture on temperature extremes

This section describes the various important factors associated with the SM-T interaction. The present study focused on six important factors (i.e. sensible heat flux: SHF, latent heat flux: LHF, evapotranspiration: ET, soil moisture: SM, maximum temperature: $T_{max}$, soil moisture memory: SMM), which extend a major contribution in understanding the linkage between SM and ExT. Three different model experiments i.e. control run (HIST), HIST-20 and HIST+20 were used here to explain the role of SM on ExT. Fig. 13 shows the spatial distribution of the long-term mean (climatology) change in SM, $T_{max}$, SHF, LHF, ET and SMM for DRY-SM (HIST-20) experiment (1st column) as well as WET-SM (HIST+20) experiment (2nd column) with respect to historical climate. Fig. 13 reveals the sensitivity of ExT to SM through the modulation of surface energy partitioning (ratio of SHF to LHF), ET, SMM and $T_{max}$. ET is one of the important factors in land-atmosphere coupling processes, which is mainly controlled by SM and energy availability at the land surface. Based on the evaporative fraction (ratio of LHF to net radiation), Seneviratne et al. (2010) classified the ET regimes as a function of surface SM i.e. i) SM bounded, ii) energy bounded, and iii) transitional climate regime. Fig. 13i, j and Fig. 1o indicate that the regions where SM conditions are found to be wetter or drier have less impact on ET variability. These regions are mainly located over the north-west, north and north-east parts of India. On the other hand, maximum sensitivity of ET can be observed over moderate SM regimes, where enough amount of SM and energy is available. Quantitatively, 20% decrease (HIST-20) of SM over the transitional climate zone of NCI can lead to the decrease of the ET by 10%, and a 20% increase (HIST+20) of SM can enhance ET by 15%.

Furthermore, limiting the total available energy for the latent heating process (i.e. energy consumed during the evapotranspiration process), SM dominantly controls the surface energy partitioning at the land surface. The effect of SM on surface energy partitioning is shown using the sensitivity experiments in Fig. 13e, f, g and h. A decrease of SM over the transitional climate regime of NCI leads to contributing more radiational energy for a sensible heating process by limiting the energy used for LHF. An increase in long-term mean SHF over the strong SM-T coupling region generally takes place due to the more amount of energy consumed for heating the atmosphere through enhanced dry and warm land surface conditions (Ganeshi et al., 2020). Thus, drier SM conditions induce more warm atmospheric conditions over the strong SM-T coupled regions. Whereas, WET-SM experimental results are indicating a relatively cool near-surface atmosphere due to entertainment of less amount of SHF through surface energy partitioning.

Soil moisture memory (SMM) is another crucial aspect of the climate system that affects land-atmosphere interactions(Ganeshi et al., 2020; Mujumdar et al., 2021; Orth and Seneviratne, 2013). Therefore, in the present study, we have evaluated the SMM characteristics over the Indian region with the help of an e-folding time scale (described in subsection 2.3.4) from the historical as well as WET-SM (HIST+20) and DRY-SM (HIST-20) sensitivity experiments. This analysis aims to highlight the sensitivity of SMM to WET-SM and DRY-SM simulations. The findings of the MRI model output suggest that the SMM time-scale over the Indian region varies from one to eight weeks (Supplementary Fig. 4). The model shows the lowest SMM time scale (< 2 week) over drier regions of central as well as north-west Indian regions. However, the highest persistence of SM time scale (> 5 weeks) is found over the north and north-east Indian regions. Furthermore, most of the regions of central India, including a strong SM-T coupling zone of NCI, indicate a moderate SM persistence time scale of about 1 to 5 weeks. A study by Delworth and Manabe (1993) has linked the SMM time-scale with the persistence of atmospheric variability and thus consecutively on near-surface temperature. In comparison to wet and dry SM regimes, moderate SM zones will experience faster evaporative damping of SM anomalies due to available radiational energy, and so have the potential to influence near-surface temperature variability. The results pointed out the decrease of the SMM time scale almost everywhere over an Indian region in the DRY-SM sensitivity experiment (Fig. 13 k). The results also highlight that the SMM time scale over the weak coupling regions will not be modified by the change in SM content due to lower sensitivity. On the other hand, SMM time-scale behaviour is highly non-linear in WET-SM experiments over the Indian region (Fig. 13 l). Model output shows that SMM can increase or decrease with the WET-SM experiment (HIST+20). Over the north-central Indian region, dry soil moisture (HIST-20) conditions lead to reducing the persistence time scale by 1 week and wet soil moisture (HIST+20) intensifies the SMM by a few days (< 1 week). In summary, drier SM conditions over the strongly coupled region will cause the SM anomalies to dissipate faster due to evaporative damping. In principle, faster dissipation of SM anomalies favours near-surface temperature warming by reducing evapotranspiration and increasing sensible heat flux across the strongly coupled region of NCI. On the other hand, further investigation need to be carried out to study the factors responsible for non-linear behavior between SMM and WET-SM over the weak coupling zone.

## 4. Summary and conclusion

The present study evaluates the impact of soil moisture (SM) variability on long-term changes in annual extreme temperature characteristics (frequency, duration and intensity) over India for historical (1951-2010) and future (2051-2100) climates. To fulfil the aim, the study performs control as well as a set of SM sensitivity experiments (WET-SM and DRY-SM) by using a high-resolution (~60 km) version of the MRI-AGCM3.2. The sensitivity experiments are initialized by decoupling the SM (20% increase or decrease of SM) on 1st day of each month from the control run (FUT and HIST), then further freely evaluating with atmospheric forcing. The study mainly highlights the role of SM on ExT over hot-spot of strong land-atmosphere coupling. Bias corrected outputs from MRI-AGCM3.2 used in the present study for analysis of ExT, land-atmosphere coupling processes and SM influence on extremes. Corrected model outputs are showing a good agreement with IMD observation and assimilation products (LDAS and GLDAS). SM-T coupling is evaluated here using the regression of $T_{max}$ anomalies on SM anomalies as suggested in the paper by Dirmeyer (2011). Furthermore, ExT diagnosis over the region of strong SM-T coupling for sensitivity experiments is also supported by Generalized Extreme Value Distribution (GEV) analysis and Probability Distribution Function (PDF) analysis for quantitative assessment of SM impact.

SM-T coupling strength in the MRI-AGCM3.2 is investigated here using the regression method. The spatial pattern of SM-T coupling indicates a hotspot located over the north-central Indian (NCI) region. The result from SM-T coupling is similar to the previous study by Ganeshi et al. (2020). This analysis pointed out that higher coupling strength over the region of NCI and indicates the dominant role of SM on temperature by controlling the surface energy partitioning. The SM-T coupling strength is further evaluated here in future climate under the 4K warming scenario. Results from the future 4K warming scenario shows a similar spatial distribution of SM-T coupling strength over India as that of historical climate with a hotspot of coupling is located over the NCI. However, it is to be noted from Fig. 3b SM-T coupling in future appears to be stronger over the NCI under the 4K warming scenario. Here, stronger coupling points towards a higher impact of SM on ExT in future over the NCI.

The analysis of ExT over India reveals an occurrence of at least 3-4 extreme events per year in the historical climate with an average duration of 5 to 6 days per event. Furthermore, the average intensity of extremes is found to be maximum over the central Indian region (> 47 $^{o}$C). Future projection using MRI-AGCM3.2 under 4K warming scenario indicates an intense extreme event (average intensity greater than historical climate) with ExTF > 10 events/year and ExTD > 15 days/event. Under the 4K warming scenario, extreme temperature cases in future climate were likely to be two times higher than that of historical climate. Extreme temperature analysis is also extended over the NCI using the area-averaged time series of ExTF, ExTD and ExTI, where the SM-T coupling is stronger. The results indicate an occurrence of at least 5 extreme events for HIST over NCI with average intensity exceeding 46$^{o}$C and a duration of 5-6 days per occurrence. Whereas, Future estimates demonstrate very intense (ExTI > 50 $^{o}$C), long persisting (ExTD > 10 days per event) and more frequent (ExTF>15 events per year) extreme events as those of historical climate under 4K warming scenario. The multiple order higher difference with respect to historical climate indicates the severity extremes in future climate. The stronger SM-T coupling strength in future is one of the contributing factors for higher ExT in the case of dry SM conditions.

The impact of SM on ExT in historical and future climates is documented here using sensitivity experiments. These simulation experiments revealed that DRY-SM conditions can dominantly increase the extreme temperature frequency and duration over India. Whereas, WET-SM conditions reduce the extreme temperature intensity over the Indian region. A relatively higher impact of SM on ExT is found over strongly coupled regions of NCI. For historical climate, a 20% decrease of SM over NCI can lead to a rise in ExT cases by ~5 events/year, ExTD by ~1.8 days per event and ExTI by ~0.71$^{o}$C. On the contrary, a 20% increase in SM can restrict the ExT over NCI below the normal condition. For future 4K experiments, DRY-SM conditions will enhance the frequency, duration and intensity of ExT by ~2.2 events/year, 1.55 days per event and 0.93$^{o}$C respectively as compared to the control run. On the other hand, WET-SM conditions can significantly reduce the extreme characteristics of the NCI. The result shows that the FUT+20 experiment decreases the ExTF by 3.3 events/year, ExTD by 2 days per event and ExTI by 2.02 $^{o}$C. In a warmer climate, precipitation and soil moisture are likely to increase in MRI-AGCM3.2 as well as the CMIP6 simulations (Moon and Ha, 2020). The wetter soil conditions are likely to act to attenuate the ExT as diagnosed in this study. Whereas, once drier SM conditions occur in a warmer climate, it exacerbates the extreme temperature conditions that lead to serious societal impacts. The present study also uses Generalized Extreme Value (GEV) Distribution analysis to estimate the role of SM on record-level ExT over the NCI. The difference between 50-year return values of ExTI from DRY-SM and WET-SM simulation for the historical climate is about 1.25$^{o}$C over the NCI. The GEV result highlights the strong influence of SM on record-level ExT over the NCI in future as compared to the historical climate. The difference between 50-year return values of ExTI from FUT-20 and FUT+20 simulations can exceed 3$^{o}$C over the NCI. The experiment also suggests an increase in SM-T coupling strength over the NCI in future climate under a 4K warming scenario. In addition, probability distribution function analysis brought out the maximum impact of SM on ExT over NCI during the post-monsoon season as compared to monsoon and pre-monsoon seasons.

Finally, the linkage between SM and ExT is shown in this study using the four important factors related to the water and energy cycle. These factors cover the evapotranspiration (ET), soil moisture memory (SMM), sensible and latent heat flux (SHF and LHF). DRY-SM and WET-SM sensitivity experiments also identified that influence of SM on ExT is closely associated with four important factors of energy and water cycle. In wet conditions, the amount of energy available at the surface is used to cool the near-surface atmosphere by increasing the LHF through the ET process. As a result, the near-surface air temperature is remaining below normal conditions, and high temperature occurrences are gradually slowed. While, in case of dry sensitivity run, below normal SM conditions cause the sensible heating process to entrain more energy back into the environment by reducing the ET rate. Drier SM conditions also diminish long term SM memory time-scale (SMM) of remembering the positive anomalies caused by the atmospheric forcing (mainly the precipitation). Consecutively, dry sensitivity experiments suggest the higher impact of soil moisture to increase the extreme temperature conditions by changing the ET, SMM and surface energy partitioning.

Acknowledgements:

The authors are grateful to the Director, Indian Institute of Tropical Meteorology (IITM, India), for the support to carry out this research work. The IITM HPC support is duly acknowledged. The IITM is fully funded by the Ministry of Earth Sciences, Government of India. The authors would like to thank Dr. Ryo Mizuta and Dr. Yukiko Imada at the Meteorological Research Institute for their support on model experiments.

Figure 1

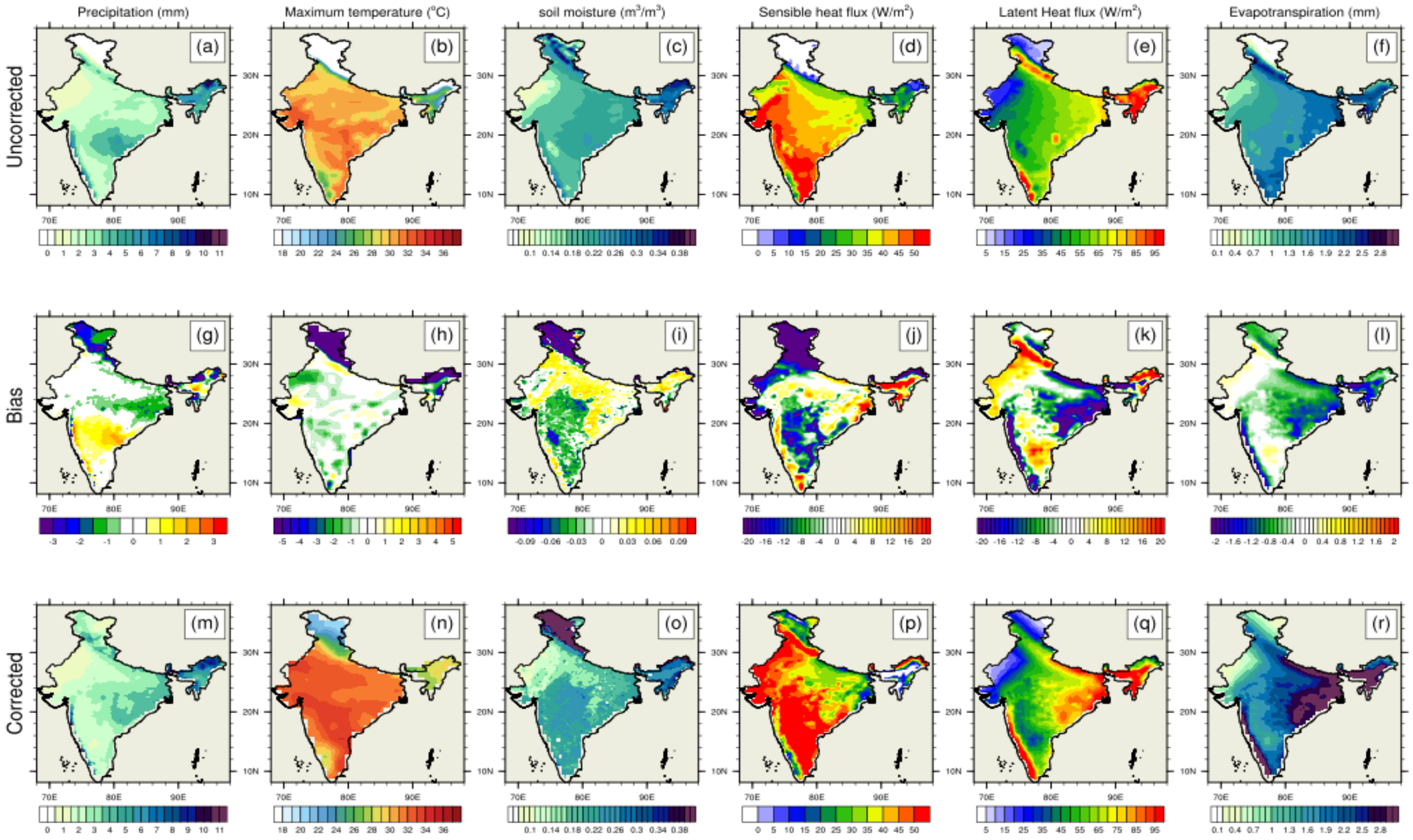

**Figure 1:** Spatial map of annual mean uncorrected (1st row), bias (2nd row) and corrected (3rd row) precipitation (PR) in mm (1st column), maximum temperature (Tmax) in °C (2nd column), soil moisture (SM) in $m^3m^{-3}$ (3rd column), sensible heat flux (SHF) in W/m$^2$ (4th column), latent heat flux (LHF) in W/m$^2$ (5th column), evapotranspiration (ET) in mm (6th column) for the historical climate (1951-2010).

**Figure2**

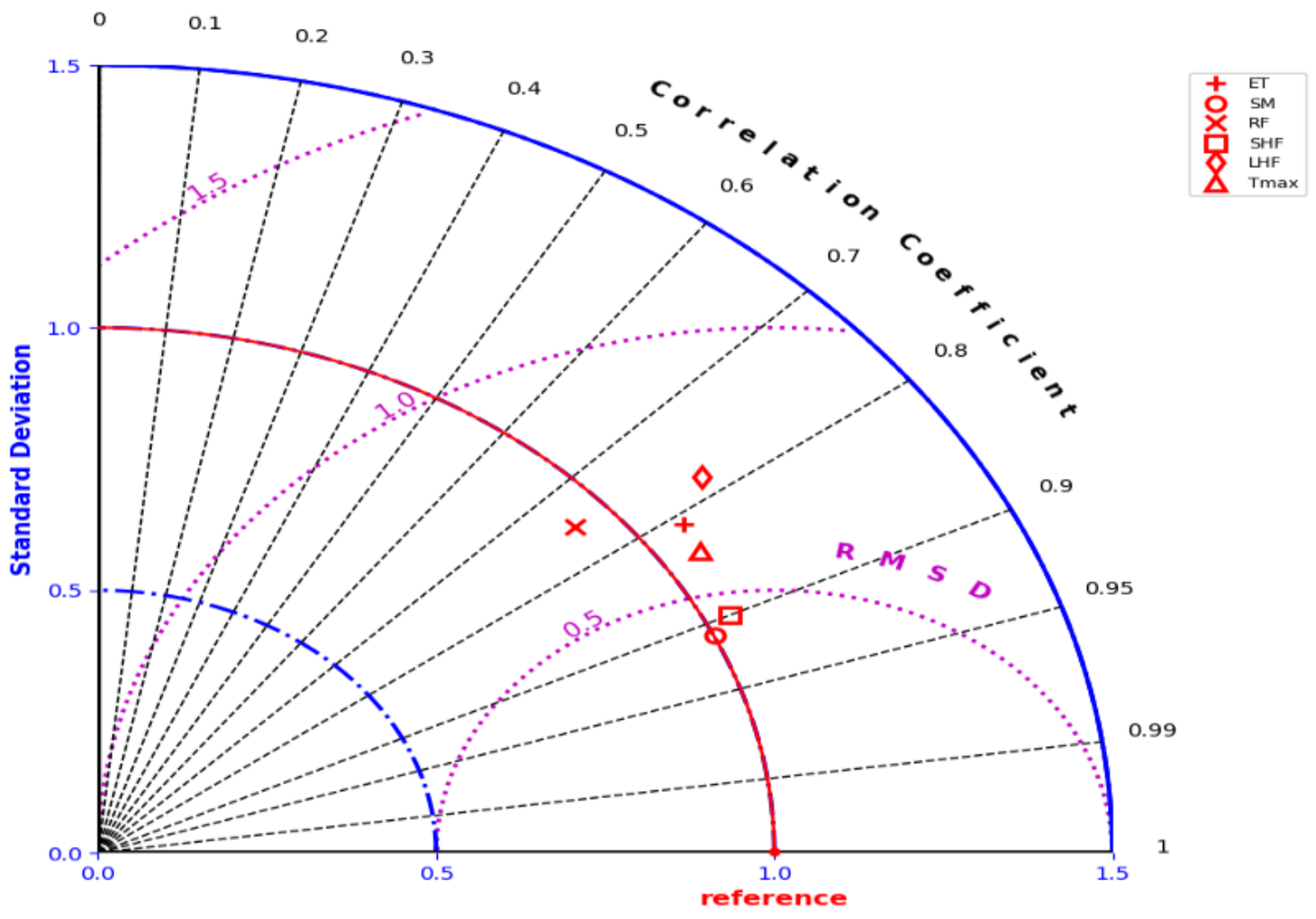

**Figure 2**: All India average Taylor statistics of PR, Tmax, SM, SHF, LHF and ET represented using the correlation coefficient, standard deviation and root mean square deviation (RMSD) with respect to validation data sources from IMD, GLDAS and LDAS. All the time series are standardized before computing the statistical scores.

Figure 3

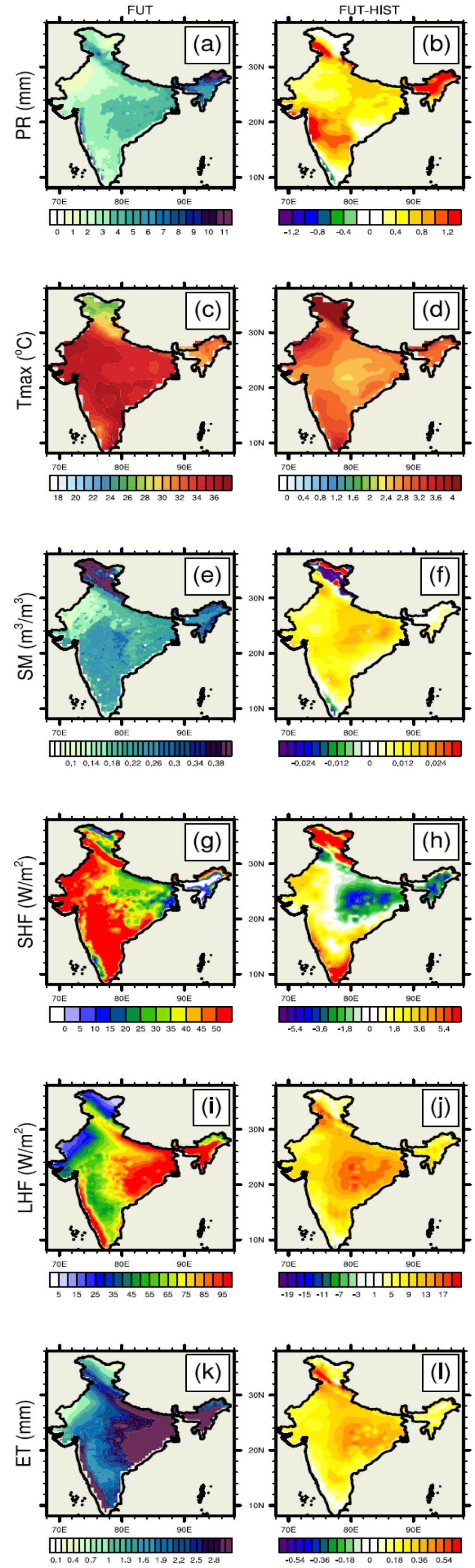

**Figure 3**: Spatial map of annual mean precipitation in mm (1st row), Tmax in °C (2nd column), SM in $m^3m^{-3}$ (3rd column), SHF in W/m² (4th column), LHF in W/m² (5th column), ET in mm (6th column) for the FUT experiment (1st row), and the difference between FUT and HIST experiment (2nd row).

Figure 4

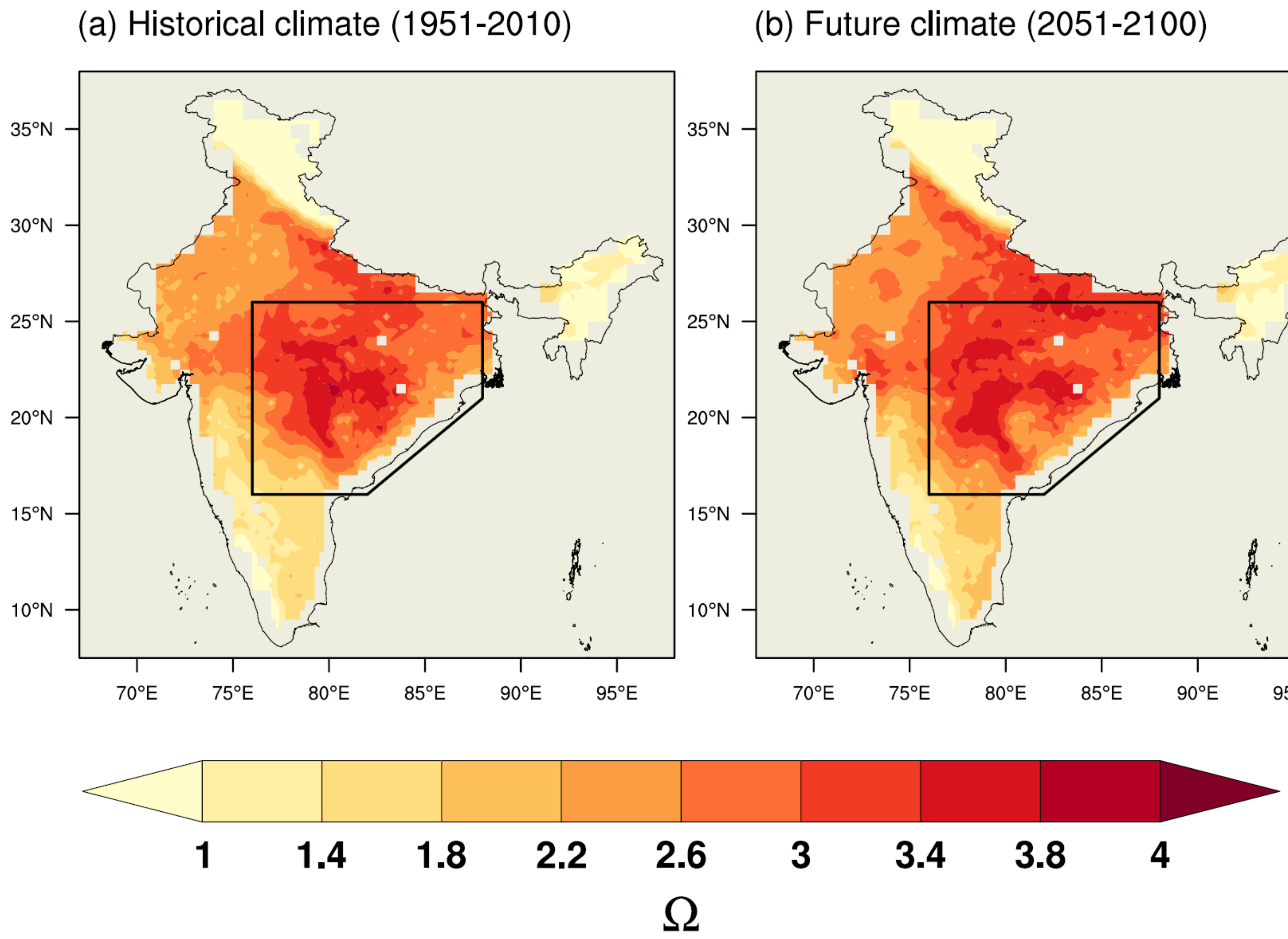

**Figure 4:** Spatial map of soil moisture-temperature coupling over the Indian region estimated using the method by Dirmeyer (2011) for the (a) HIST and (b) FUT experiments. The area shown in polygon over the north-central India (NCI) is highlighted as the region of strong soil moisture-temperature coupling over the India (75°E-87°E, 16°N-26°N, land only).

Figure 5

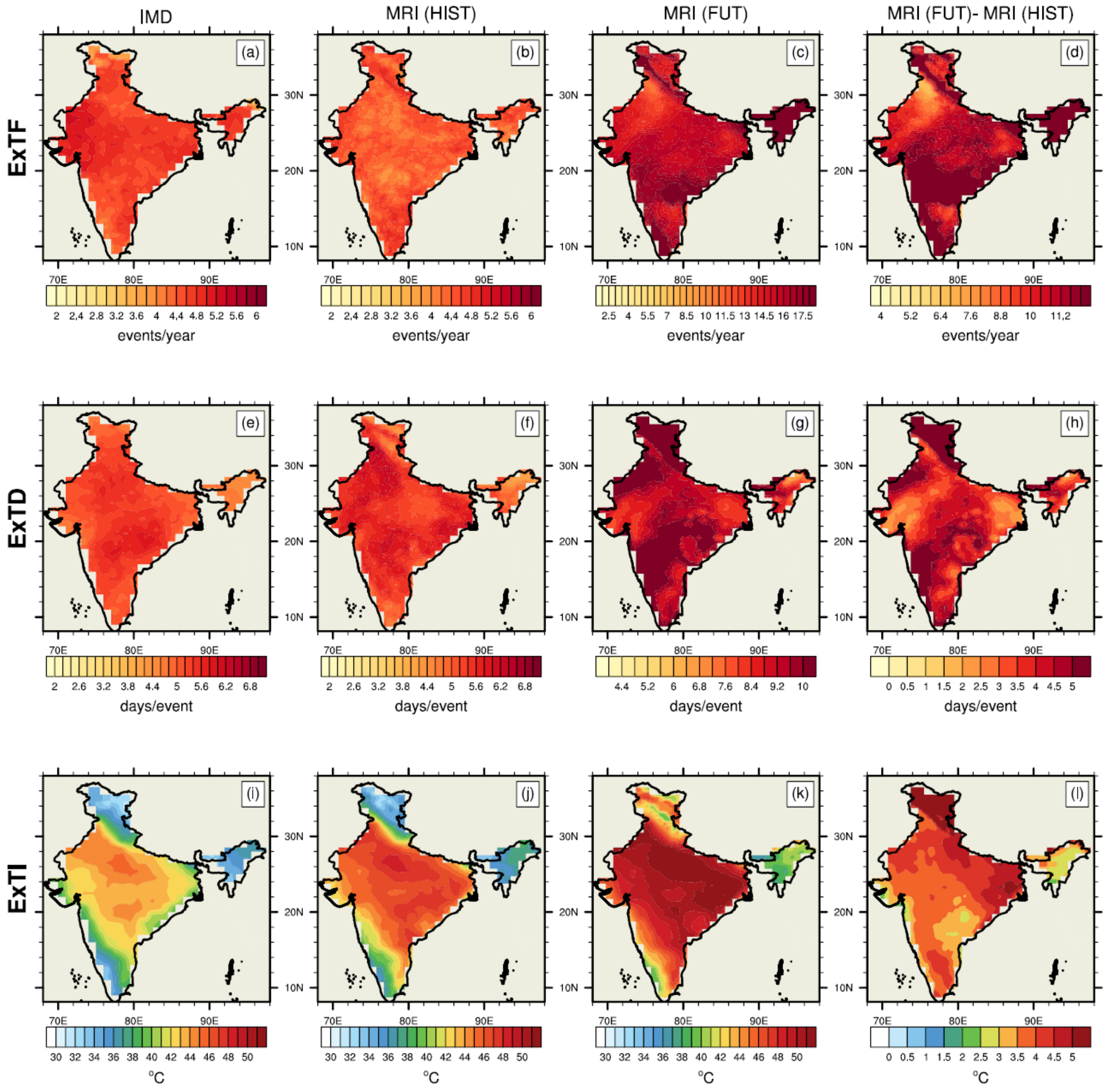

**Figure 5**: Climatology of extreme temperature frequency (ExTF), duration (ExTD) and Intensity (ExTI) for IMD (Historical) data (1st column), HIST experiment (2nd column), FUT experiment (3rd column), and difference between FUT and HIST experiment (4th column).

Figure 6

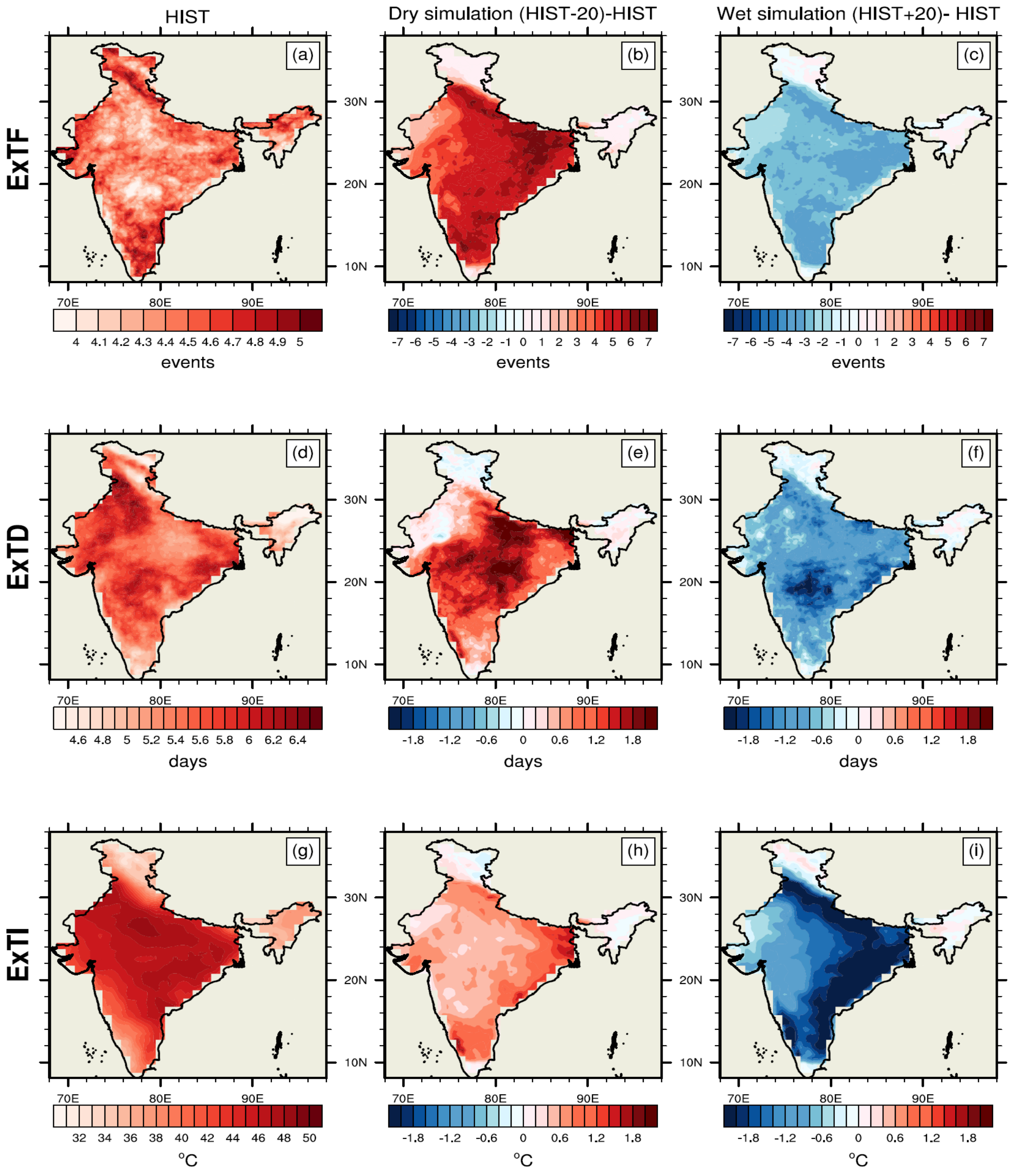

**Figure 6**: **Impact of SM on ExT during historical climate**: Climatology of ExTF (1st row), ExTD (2nd row) and ExTI (3rd row) for HIST (1st column), difference between HIST-20 and HIST experiment (2nd column), difference between HIST+20 and HIST experiment (3rd column) during the historical period 1951-2010.

Figure 7

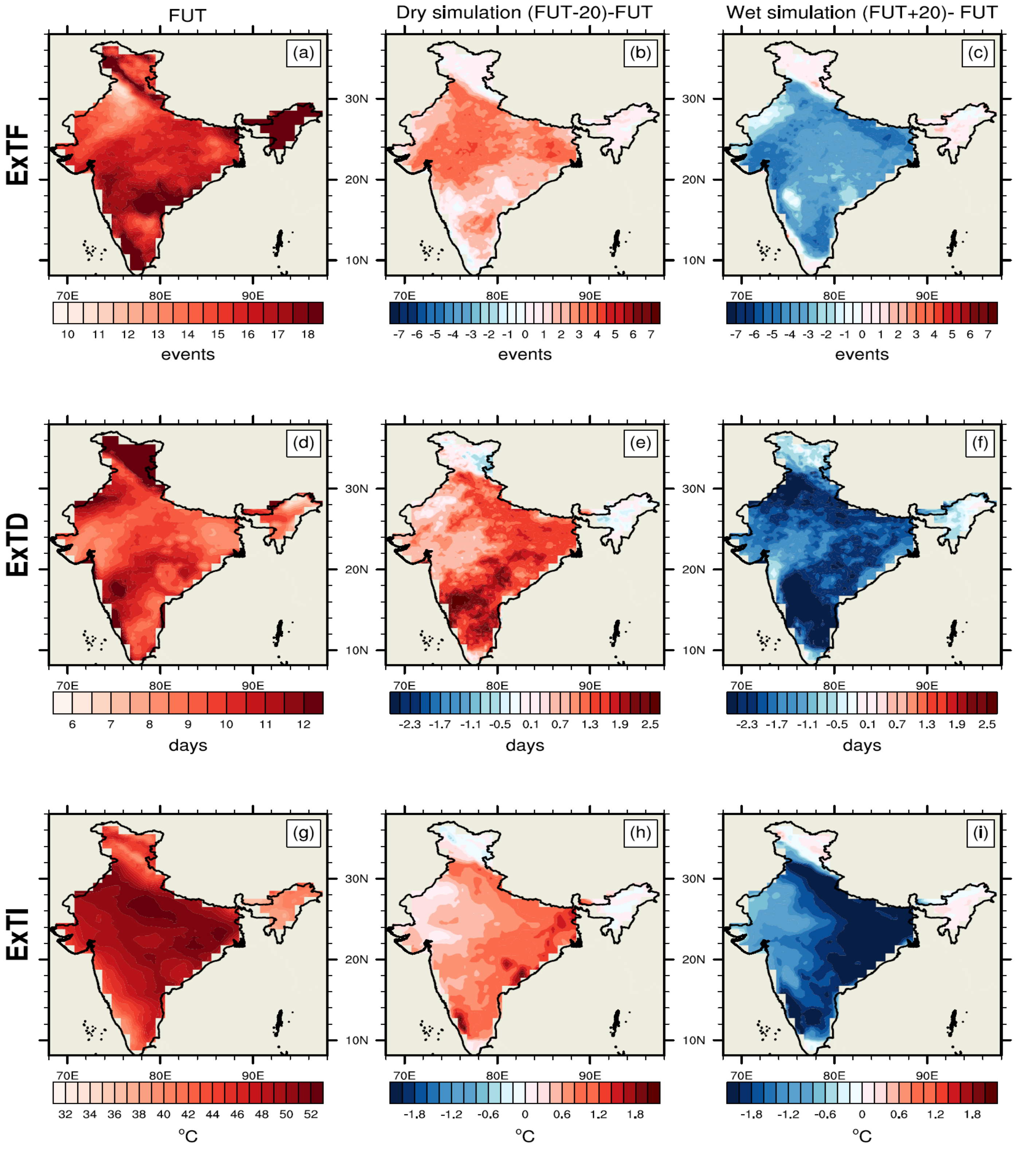

**Figure 7**: **Impact of SM on ExT during future climate**: Climatology of ExTF (1st row), ExTD (2nd row) and ExTI (3rd row) for FUT (1st column), difference between FUT-20 and FUT experiment (2nd column), difference between FUT+20 and FUT experiment (3rd column) during the future climate 2051-2100.

Figure 8

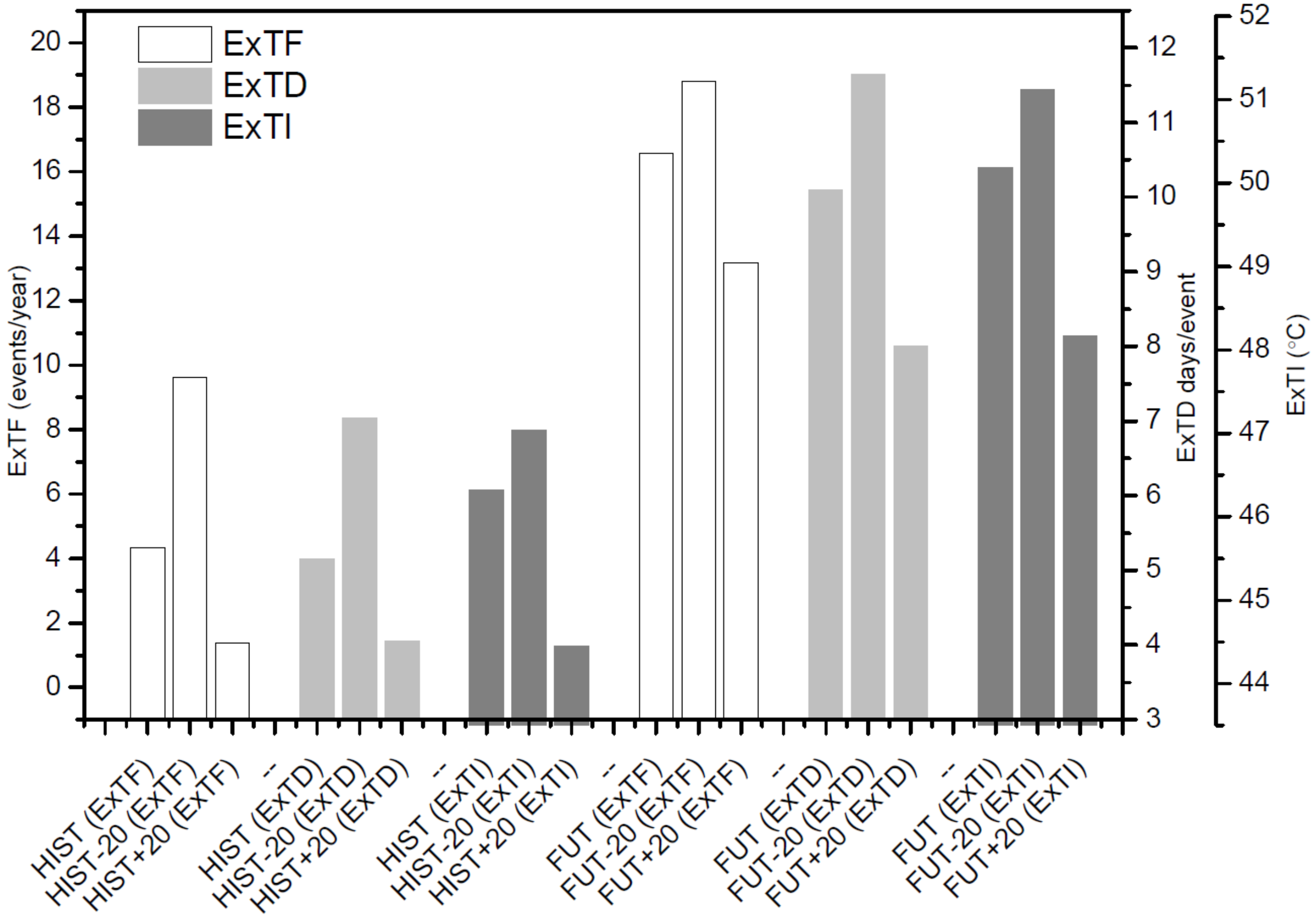

**Figure 8**: **Impact of SM on ExT over the NCI during the historical and future climate:** Histogram compares the climatology of ExTF (white), ExTD (light grey) and ExTI (dark grey) for six experiments i.e. HIST, HIST-20, HIST+20, FUT, FUT-20 and FUT+20, averaged over the NCI (75°E-87°E, 16°N-26°N, land only).

Figure 9

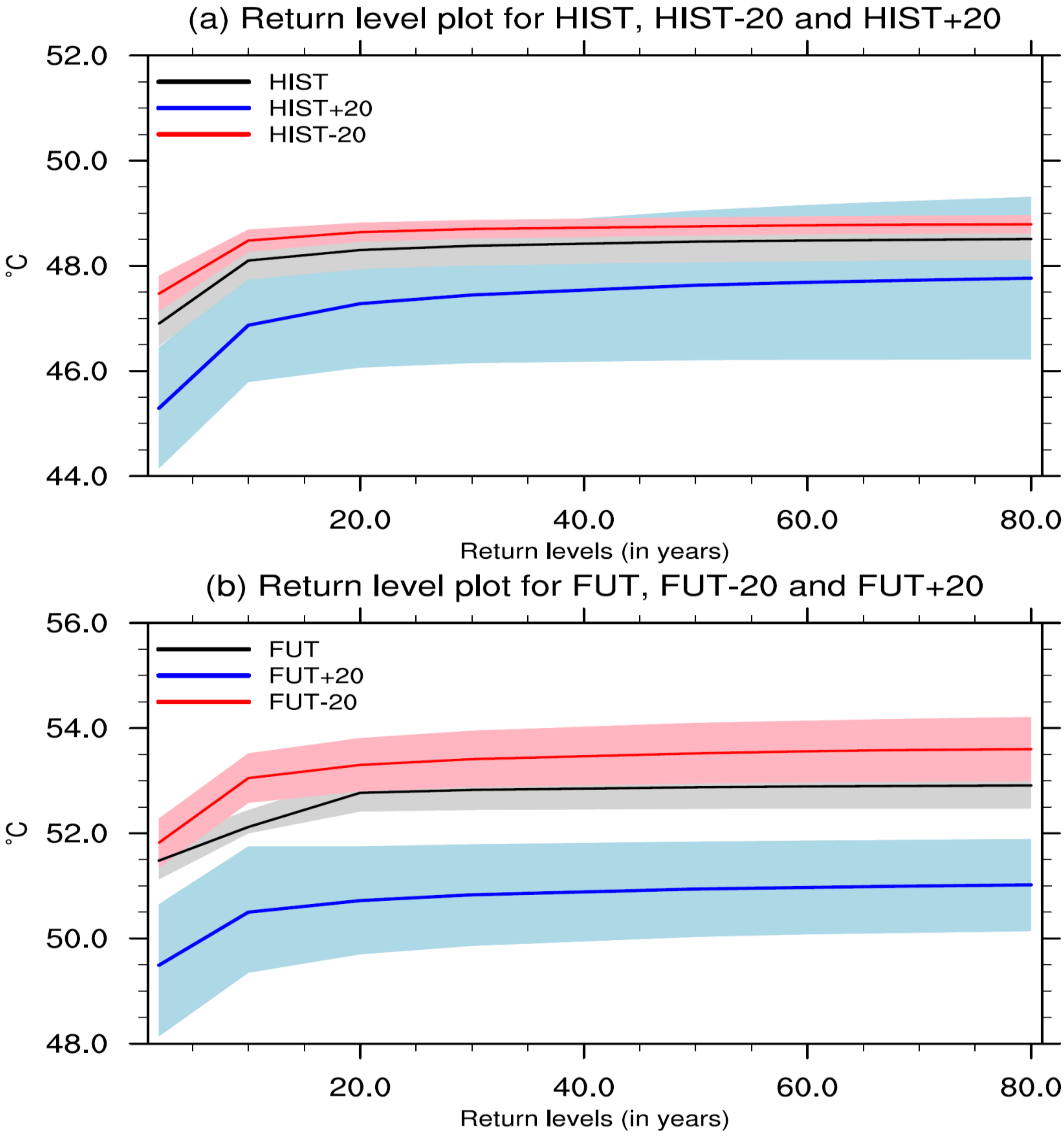

**Figure 9: Generalize extreme value (GEV) distribution:** Return level plot of ExTI for control run (black line), DRY-SM experiment (red line) and WET-SM experiment (blue line) over NCI estimated using non-stationary GEV model for (a) historical and (b) future climate. The area between upper and lower confidence interval of the return levels for control, DRY-SM and WET-SM are filled with the light grey, light red and light pink color, respectively.

Figure 10

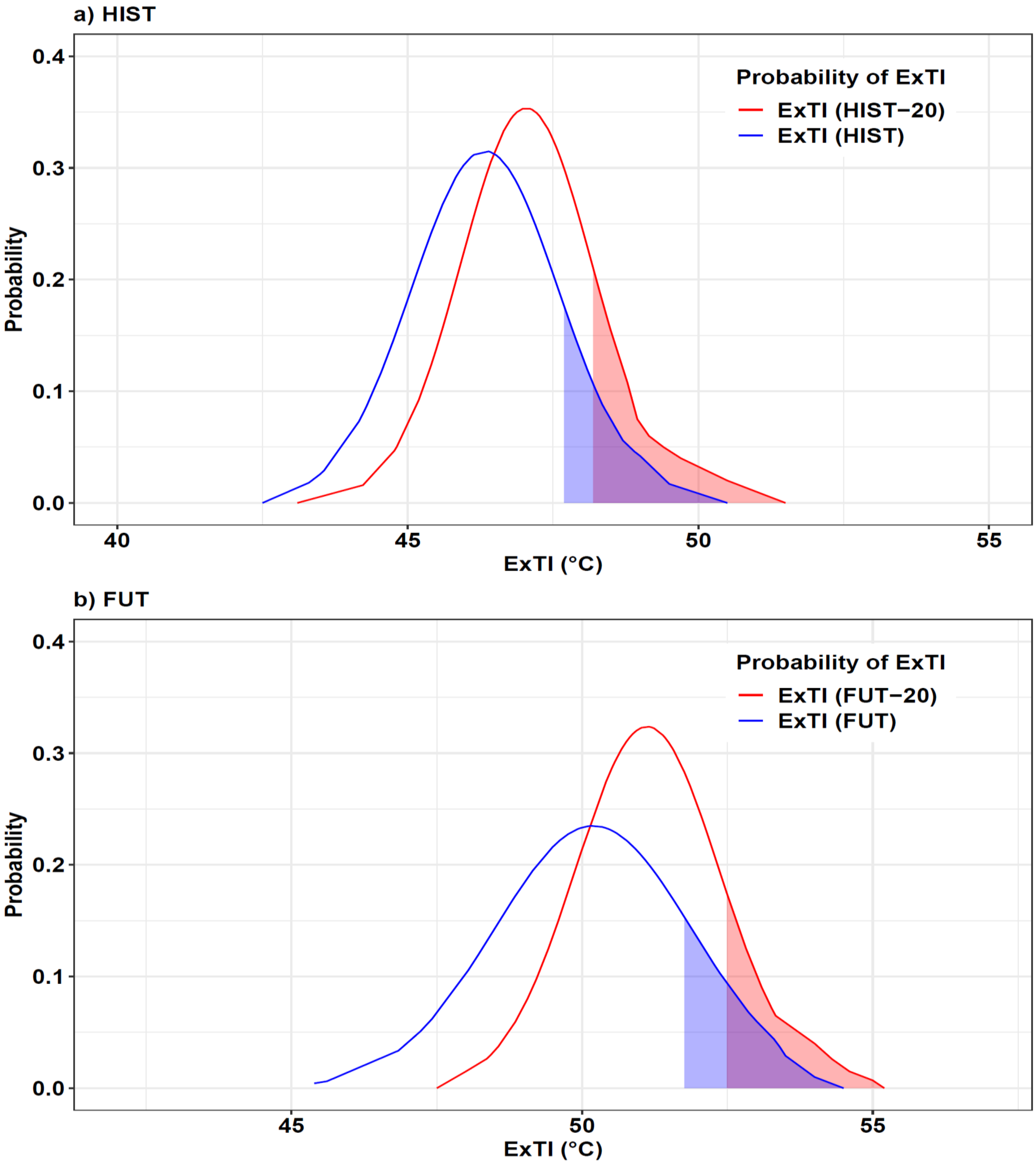

**Figure 10:** Probability density function of ExTI over the NCI for control (blue line) and DRY-SM (red line) experiments during the (a) historical and (b) future climate. The regions above the 90$^{th}$ percentile quantile of ExTI are shaded.

Figure 11

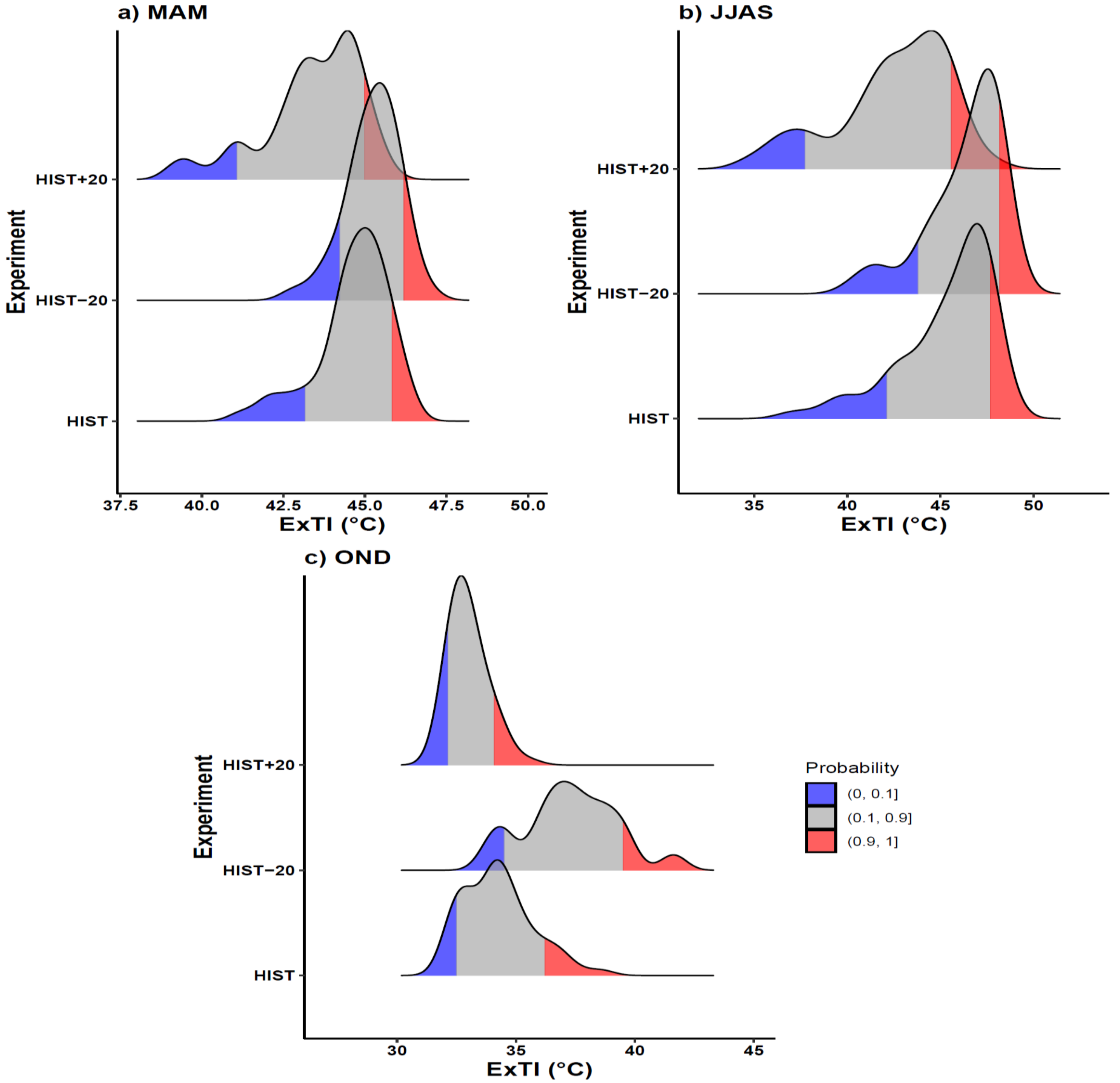

**Figure 11: Impact of SM on ExT over the NCI during historical climate for three seasons:** Comparison of Probability density function of ExTI over the NCI for HIST and HIST-20 and HIST+20 experiments during the (a) pre-monsoon (March to May: MAM) and (b) monsoon (June to September: JJAS) and, (c) post-monsoon (October to December: OND) seasons. The regions below 10$^{th}$ percentile quantile and above the 90$^{th}$ percentile quantile of ExTI are shaded with blue and red color, respectively.

Figure 12

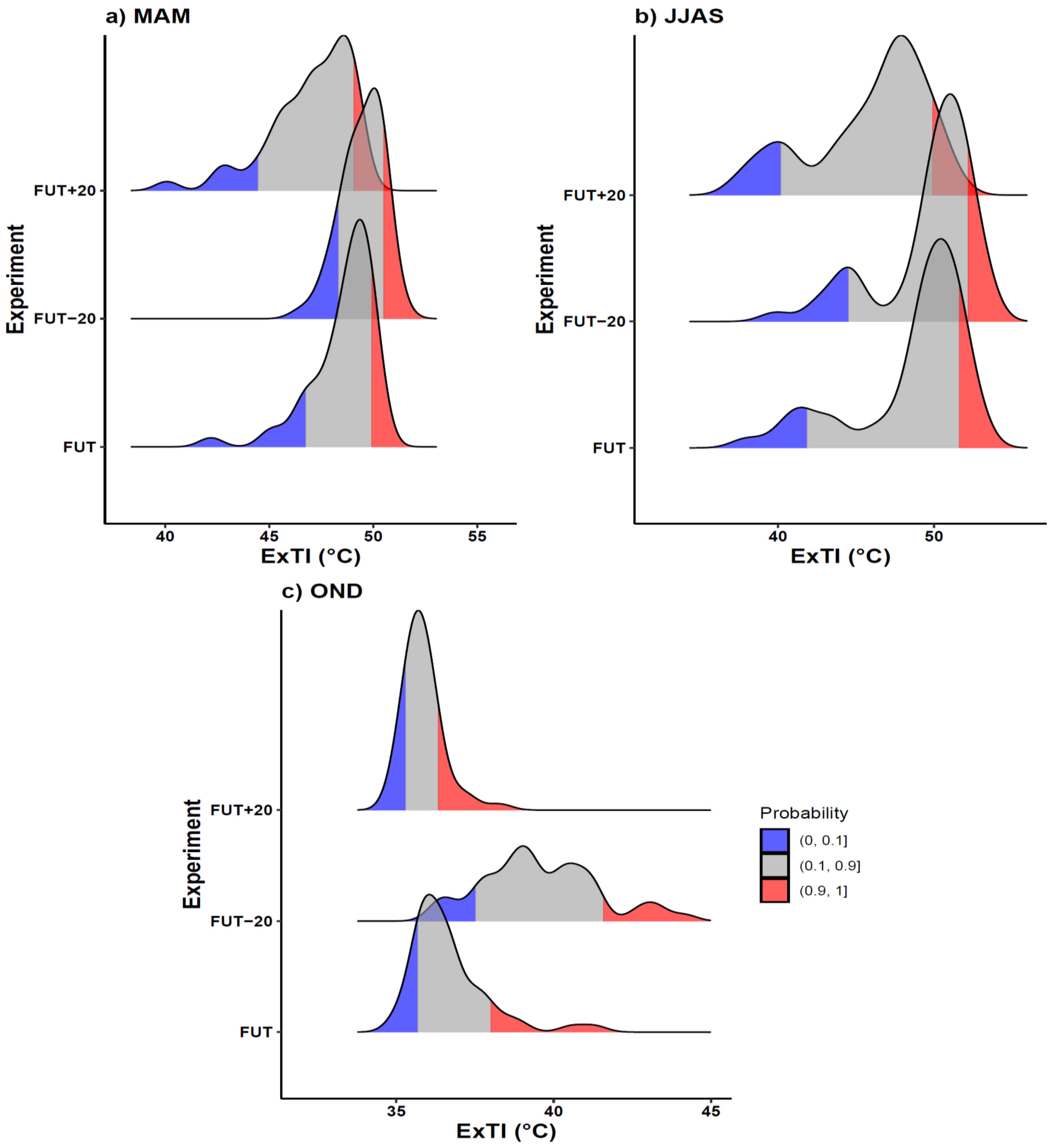

**Figure 12: Impact of SM on ExT over the NCI during future climate for three seasons:** Comparison of Probability density function of ExTI over the NCI for FUT and FUT-20 and FUT+20 experiments during the (a) pre-monsoon (March to May: MAM) and (b) monsoon (June to September: JJAS) and, (c) post-monsoon (October to December: OND) seasons. The regions below 10$^{th}$ percentile quantile and above the 90$^{th}$ percentile quantile of ExTI are shaded with blue and red color, respectively.

Figure 13

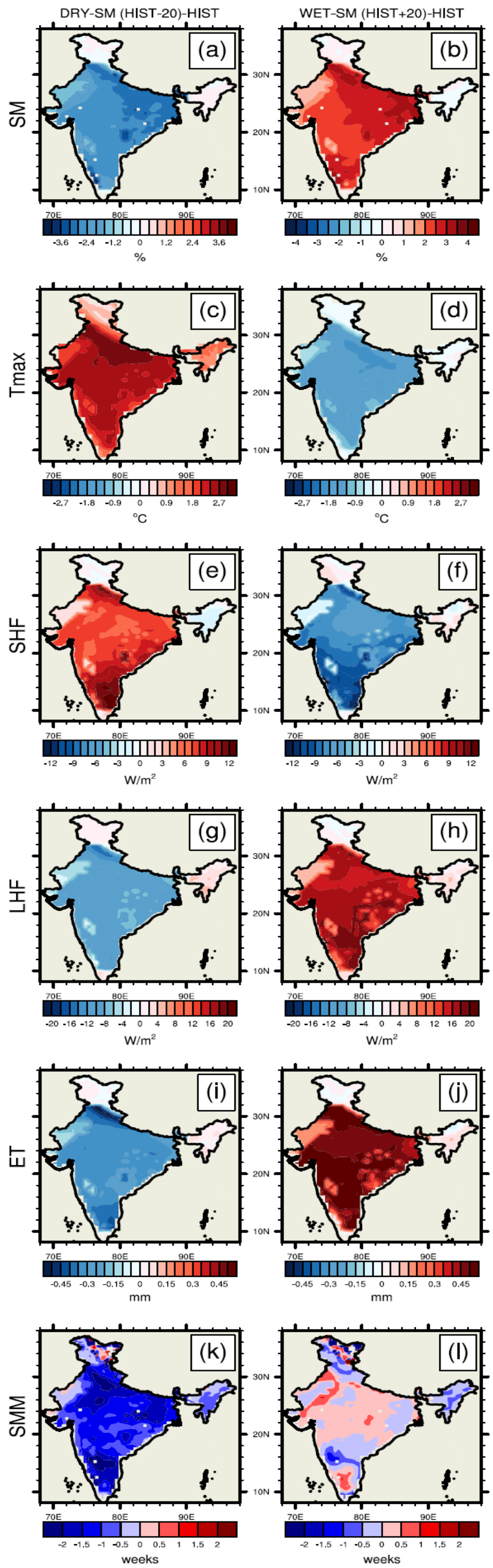

**Figure 13: Factors contributing the land-atmosphere interaction:** Spatial maps of difference between annual mean SM (1st row), Tmax (2nd row), SHF (3rd row), LHF (4th row), ET (5th row) and SMM (6th row) from (a) DRY-SM (HIST-20) and HIST (1st column), (b) WET-SM (HIST+20) and HIST (2nd column).

Supplementary figure S1

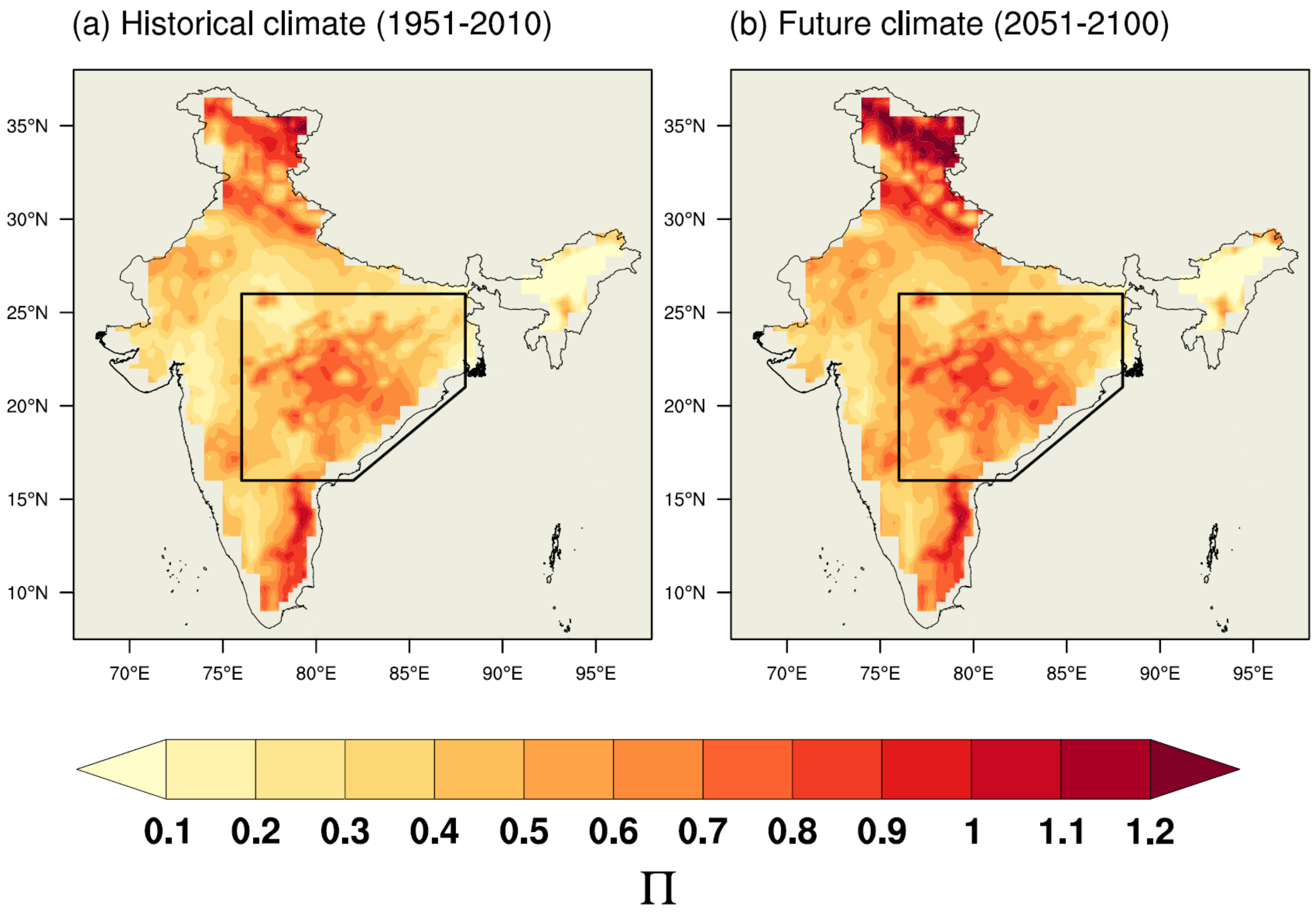

Spatial map of soil moisture-temperature coupling over the Indian region estimated using the method by Miralles et al. (2012) for the (a) HIST and (b) FUT experiments. The area shown in polygon over the north-central India (NCI) is highlighted as the region of strong soil moisture-temperature coupling over the India (75°E-87°E, 16°N-26°N, land only).

Supplementary figure S2

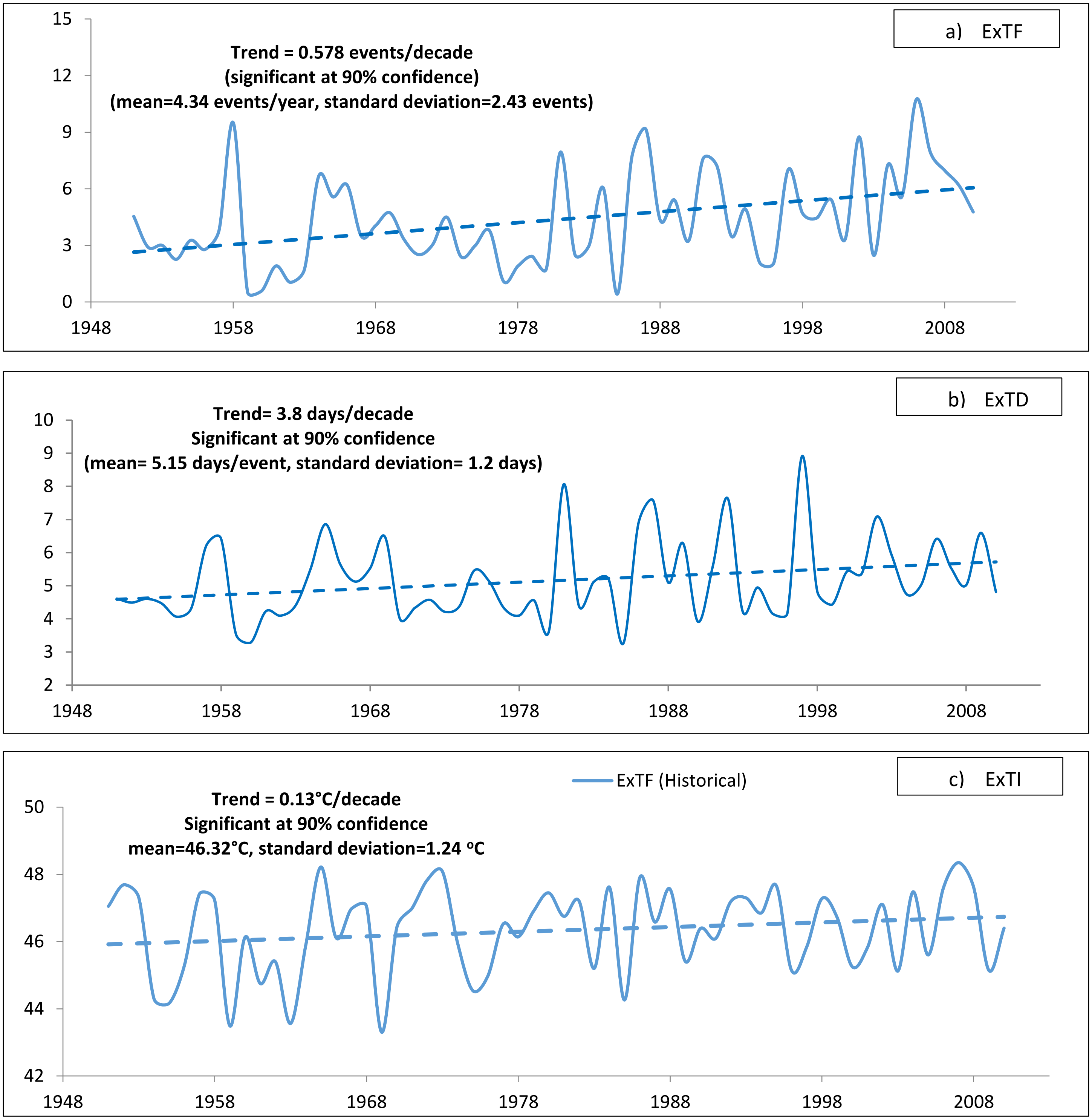

Time series of (a) annual count of extreme temperature frequency (ExTF), (b) annual count of extreme temperature duration (ExTD) and (c) extreme temperature intensity (ExTI) averaged over the NCI (75°E–87°E,16°N–26°N, land only) for historical period (1951-2010).

Supplementary figure S3

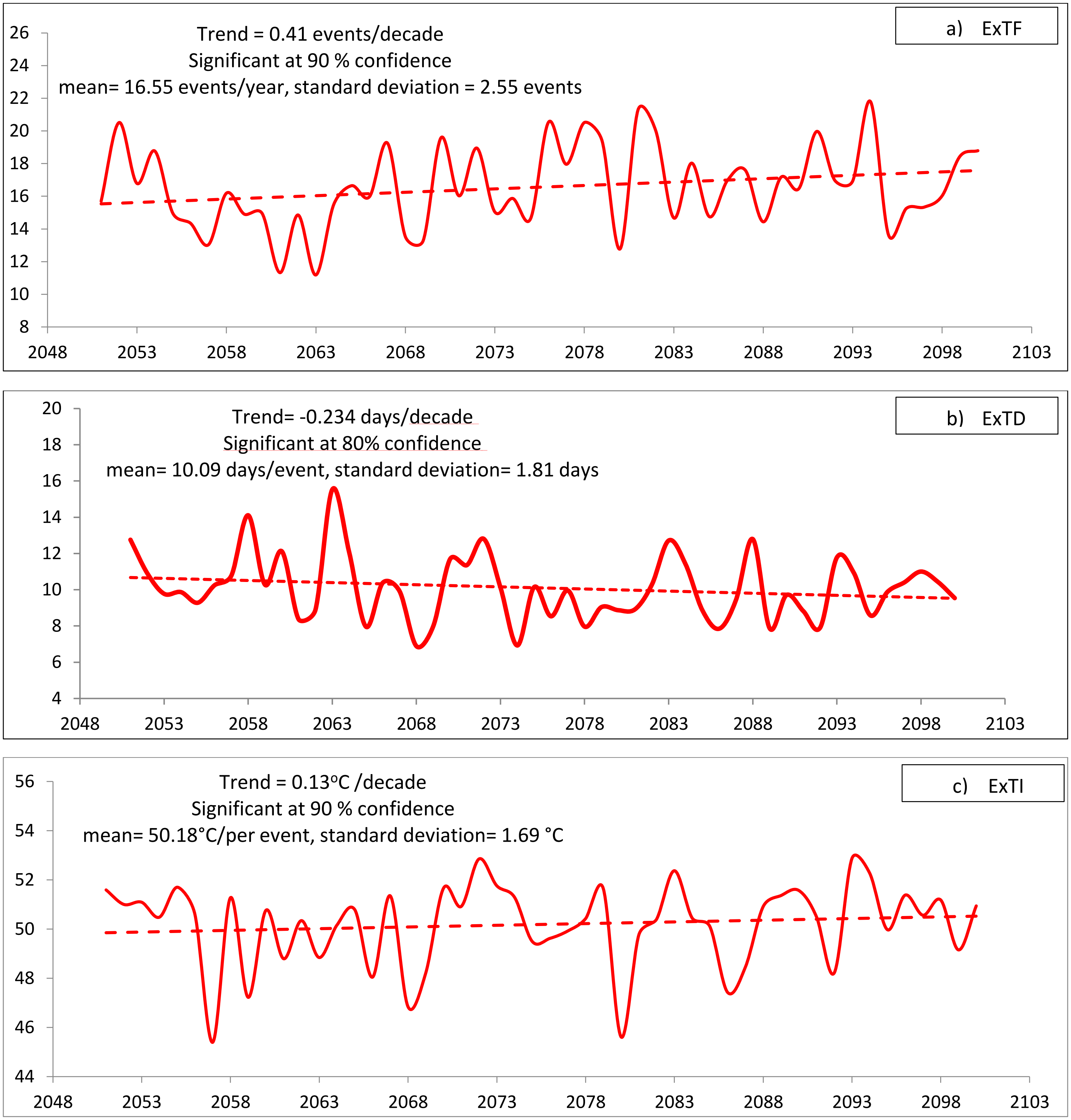

Time series of (a) annual count of extreme temperature frequency (ExTF), (b) extreme temperature duration (ExTD) and (c) extreme temperature intensity (ExTI) averaged over the NCI (75°E–87°E,16°N–26°N, land only) for FUT (2051- 2100).